\documentclass[
 reprint,
nofootinbib,
 amsmath,amssymb,
 aps,
]{revtex4-2}

\usepackage{graphicx, color}
\usepackage{dcolumn}
\usepackage{bm}
\usepackage[mathlines]{lineno}
\usepackage{amsmath}
\usepackage{amssymb}
\usepackage{float}
\usepackage{mathrsfs}
\usepackage{lipsum}
\usepackage{hyperref}
\usepackage[normalem]{ulem}
\hypersetup{
    colorlinks=true,       
    linkcolor=red,          
    citecolor=rp,        
    filecolor=magenta,      
    urlcolor=blue           
}
\usepackage{cleveref}
\usepackage{color}

\newcommand{\mpl}{m_\mathrm{pl}}

\definecolor{rp}{cmyk}{0.2, 1, 0.6, 0}
\definecolor{rp}{cmyk}{0.2, 1, 0.6, 0}
\definecolor{green2}{cmyk}{0.27, 0, 1, 0.52}

\newcommand{\bPsi}{{\bm{\mathsf{\Psi}}}}
\newcommand{\bepsilon}{{\bm{\mathsf{\epsilon}}}}
\newcommand{\bS}{{\bm{S}}}

\newcommand{\bR}{{\bm{\mathsf{R}}}}
\DeclareMathOperator{\Tr}{Tr}

\newcommand{\ym}{\mathrm{ym}}

\def\mpl{m_\mathrm{pl}}

\usepackage{hyperref}
\hypersetup{%
  colorlinks = true,
  linkcolor  = rp
}

\begin{document}

\preprint{APS/123-QED}

\title{Yang-Mills stars in Higgsed non-Abelian dark matter}

\author{Mudit Jain}
\email{mudit.jain@rice.edu}
\affiliation{%
Department of Physics and Astronomy, Rice University\\
Houston, TX, 77005, U.S.A.
}

\date{\today}

\begin{abstract}
Bosonic field theories with self interactions alongside gravity, generally admit bound states known as solitons. Depending upon the spin nature of the field, they can even carry macroscopic intrinsic spin polarization. Focusing on the SU($2$) case, we describe polarized solitons in non-Abelian theories with a heavy Higgs, which we refer to as `Yang-Mills stars'. Owing to both kinds of self-interactions; repulsive ones arising due to the Yang-Mills structure, while attractive ones arising due to the Higgs exchange;
we can have a diverse zoo of solitons. Depending upon various parameters of the theory such as the mass of the Yang-Mills vector fields $m$, mass of the dark Higgs field $M_{\varphi}$, and the gauge coupling constant $g$, these objects can be astrophysically large with varying size and mass, and carry large intrinsic spin and/or iso-spin giving rise to interesting phenomenological implications. Even for vector mass as large as $m \simeq 10$ eV, we can accommodate gauge couplings $g \lesssim 10^{-4}-10^{-5}$, still evading Bullet cluster constraints. For these parameters, there may exist cosmologically long lived solitons having radii as large as $r_{s} \sim 10^{5}\,R_{\odot}$ and masses $M_{s} \sim 10 M_{\odot}$, carrying $M_{s}/m \sim 10^{66}$ amounts of intrinsic spin and iso-spin polarization. As a subset of the space of soliton solutions in the SU($2$) Higgs model, in the end we also explicitly discuss solitons in the Abelian Higgs model. 
\end{abstract}

\maketitle

\tableofcontents

\section{Introduction}
\label{sec:intro}
The Standard Model (SM) of particle physics and General Relativity (GR) have, time and again, withstood all the observational tests to date. Astrophysical tests however reveal the presence of new physics beyond the SM, in particular dark matter (DM) constituting $\sim 84\%$ of the total matter in the observable Universe. Even though it primarily interacts with the SM through gravity (and only very weakly, if also through any direct couplings)~\cite{Planck:2018vyg}, there is only little understanding of the nature of particle(s)/field(s) that constitute DM. If all, or at-least most of the observed DM comprise of a single type of quanta, its mass can range anywhere between $10^{-21}\,\mathrm{eV} \lesssim m \lesssim \mpl$ ~\cite{Irsic:2017yje}, with bounds softening further in the case of composite dark matter~\cite{Jacobs:2014yca}. There is also no understanding of the spin nature of the constituent DM particles.\\

In the course of its cosmological evolution, a massive field may undergo a transition if, and when, the typical/ambient energy scale drops much below its mass $m$. This can happen, for instance, when $T \ll m$ where $T$ is the temperature of the field; or when the Hubble rate $H \ll m$. What we have at our disposal is a non-relativistic (NR) effective field theory resulting in a Schr\"{o}dinger like wave equation for the field (see~\cite{Salehian:2021khb} for a detailed analysis for spin-0, ~\cite{Adshead:2021kvl} for spin-1, and~\cite{Jain:2021pnk} for a spin-s bosonic field in general). For bosonic fields, the typical occupation numbers within the de Broglie length scale can be very large, resulting in coherent field configurations and thus warranting a classical wave description~\cite{Hertzberg:2016tal}. This is especially relevant to the case of the cold dark matter scenario. See~\cite{Bertone:2016nfn} and references therein, for a broad review. Depending upon the mass of the field, this coherent wave dynamics can manifest itself on astrophysical scales, resulting in interesting and rich phenomenology such as suppression of structure on small scales \cite{Matos:2000ss,Hu:2000ke}, turbulence \cite{Mocz:2017wlg}, vortices, interference patterns \cite{Schive:2014dra}, \cite{Hui:2020hbq}, superradiance \cite{Brito:2015oca}, solitons \cite{Schive:2014dra, Amin:2019ums,Arvanitaki:2019rax} etc, all studied in the context of spin-0 dark matter. Also see~\cite{Hui:2021tkt,Marsh:2015wka,Lee:1991ax, Liebling:2012fv, Nugaev:2019vru} for reviews and references therein. Owing to the unknown nature of the spin of the dark matter particle(s), recently there has been growing interest in such phenomenological avenues for higher (than zero) spin fields as well. For instance see~\cite{Aoki:2017ixz,Adshead:2021kvl,Jain:2021pnk,Zhang:2021xxa} for solitons in higher spin fields. In particular, there can arise polarized solitons carrying huge intrinsic spin angular momentum~\cite{Jain:2021pnk,Zhang:2021xxa}, with rich phenomenological implications owing to this spin polarization.
Also very recently, formation and implication of the vector nature of dark matter on small scale structure was explored in~\cite{Gorghetto:2022sue,Amin:2022pzv}.\\

From a theoretical point of view, top-down constructions like string theory suggest that there should be a huge dark sector (or hidden sector) having many gauge groups and various new degrees of freedom (dof)~\cite{Giedt:2000bi,Arvanitaki:2009fg,Taylor:2015ppa}. In the light of this possibility of a diverse dark sector with different gauge groups, in this paper we will explore the case of a `Higgsed' Yang-Mills (HYM) sector with a heavier Higgs (than the vector dof) to be the dark matter candidate in our Universe. We use the term `Higgsed' to reflect the presence of a (dark) Higgs field in the fundamental representation of the non-Abelian group. For simplicity we will assume no other matter fields charged under the dark SU($n$), and the theory to be in the unconfined phase at the Higgs scale, in order for the relevant dof to be the Yang-Mills vector fields at smaller energy scales. Then, integrating out the Higgs gives an effective theory for these vector dof, which admits a diverse set of non-topological solitons in the Higgs phase, that can carry macroscopically large intrinsic spin and/or iso-spin.\\

The structure, along with a brief summary of the paper, is as follows. In Sec.~\ref{Sec:effective_theory} we begin with a Higgsed non-Abelian theory and integrate out the Higgs field to get an effective theory for the vector fields. For explicitness and simplicity in this paper, we shall focus on the specific case of SU($2$) with Higgs in the fundamental representation.

Once the ambient energy scale in the dark sector falls much below the mass of the vector fields, a non-relativistic description of the non-Abelian vector fields becomes warranted. We derive this non-relativistic effective theory in Sec.~\ref{Sec:NR_effective}, and discuss the various conserved quantities in it. We also work out a constraint on the gauge coupling constant $g$, due to Bullet cluster observations. 

In Sec~\ref{Sec:grav_bound} we finally discuss the various possible solitons admitted in the SU($2$) HYM theory. The Higgs induced interaction between the vector fields is attractive, while the inherent self interactions due to the non-Abelian nature of the Yang-Mills theory are repulsive. We will see that the HYM theory admits both types of soliton solutions, i.e. either hosting attractive or repulsive self interactions within them, alongside gravity. These solitons can not only have intrinsic spin polarization, but also iso-spin owing to the non-Abelian structure. In this way, this is a natural extension of our earlier work on the spin aspect of non-topological solitons~\cite{Jain:2021pnk,Zhang:2021xxa}. We also provide estimates for lifetimes of these solitons against perturbative decays, and discuss their stability against possible parametric resonance. Note that these are not the non-Abelian Proca stars~\cite{Dzhunushaliev:2021tlm,Dzhunushaliev:2022apb} where the Higgs field settles to its vacuum value only far outside of the star, with non-trivial behavior in the interior. Neither are these objects gravitating monopoles or sphalerons~\cite{Brodbeck:1997sa}, nor glueballs (lowest energy massive states in the confined phase of pure Yang-Mills). To our knowledge, such objects have not been studied in the literature.

The case of the U($1$) Abelian Higgs model comes as a special case of the Yang-Mills model, which we highlight in Sec.~\ref{sec:Abelian.special.case}. We end by summarizing in Sec.~\ref{Sec:sum_disc}, and also discuss some possible future directions. For the purposes in this paper, we remain agnostic towards cosmological production mechanisms of such a HYM dark sector. This will be the subject of a companion paper~\cite{Jain:2022}.\\

\noindent{\bf Conventions}: Throughout this work, we adopt the following conventions: (1) we work in natural units where $\hbar = c = 1$, and adopt mostly negative signature for the metric; (2) Greek indices such as $\mu,\nu$, are used to represent space-time components while English indices such as $a,b,c$, are used to represent the internal iso-spin components; (3) a summation over repeated indices is implied, unless otherwise mentioned; (4) dot product is to denote vector dot product between spatial vectors.

\section{Effective theory}\label{Sec:effective_theory}

We begin with the Lagrangian density $\mathcal{L}_{\mathrm{full}} = \mathcal{L}_{\mathrm{hym}} + \mathcal{L}_{\rm{gr}} + \mathcal{L}_{\rm{vis}} + \Delta\mathcal{L} + ...$, where the Higgsed Yang-Mills sector is the following
\begin{align}
    \mathcal{L}_{\mathrm{hym}} =& -\frac{1}{2}\Tr[G_{\mu\nu}G^{\mu\nu}] + (D^{\mu}\phi)^{\dagger} D_{\mu}\phi\nonumber\\
    &\qquad\qquad\qquad - \lambda\left(\phi^{\dagger}\cdot\phi - \frac{v^2}{2}\right)^2\,.
\end{align}
Here, $G_{\mu\nu} = \partial_\mu W_\nu - \partial_\mu W_\nu + ig[W_\mu,W_\nu]$, $W_{\mu} = t^aW^a_{\mu}/2$ with $t^a$ being the $n^2-1$ generators of the SU($n$) algebra\footnote{The Lie algebra is $[t^a/2,t^b/2] = if^{abc}\,t^c/2$ where $f^{abc}$ are the totally anti-symmetric structure constants, and $\Tr[t^at^b] = 2\delta^{ab}$.}, and $D_{\mu} = \partial_{\mu} + ig\,W_{\mu}$ is the covariant derivative. The field $\phi$ is the dark Higgs multiplet in the fundamental representation of the SU($n$) group (a set of $2n$ real scalars), with a quartic self coupling $\lambda$ and a vacuum expectation value (vev) $v$. $\mathcal{L}_{\rm{gr}}$ and $\mathcal{L}_{\rm{vis}}$ are the Einstein-Hilbert and the visible sector (Standard Model) Lagrangian densities respectively, while the ``$...$'' represent any other possible dark sector, including the inflaton. The term $\Delta \mathcal{L}$ represents couplings between the two sectors. Writing out the kinetic term explicitly
\begin{align}
    -\frac{1}{2}\Tr[G_{\mu\nu}G^{\mu\nu}] = &-\frac{1}{4}\Bigl[F^a_{\mu\nu}F_a^{\mu\nu} - 2\,g\,f^{abc}\,F_a^{\mu\nu}\,W^b_{\mu}\,W^c_{\nu}\nonumber\\
    &\qquad + g^2f^{abc}f_{ade}W^b_\mu W^c_\nu W^\mu_d W^\nu_e\Bigr]
\end{align}
where $F^a_{\mu\nu} \equiv \partial_\mu W^a_\nu - \partial_\mu W^a_\nu$, lets us define the Yang-Mills potential
\begin{align}\label{eq:Vhym}
    V_{\rm{ym}} \equiv -\frac{g}{2}\,f^{abc}\,F_a^{\mu\nu}\,W^b_{\mu}\,W^c_{\nu} + \frac{g^2}{4}\,f^{abc}f_{ade}W^b_\mu W^c_\nu W^\mu_d W^\nu_e\,.
\end{align}
Furthermore, in the Euler representation and Higgs phase, we can set
\begin{align}
    \phi = 
    \frac{v+\varphi}{\sqrt{2}}U\hat{n}
\end{align}
without any loss of generality. Here $\hat{n}$ is an arbitrary unit (real) vector in the $n$-dimensional internal iso-space, $U(x) \in$ SU($n$), and $\varphi$ is the Higgs field (radial degree of freedom around the vev $v$). To work in Unitary gauge, we set $U^{\dagger}\,W_\mu\,U - (i/g)\,U^{\dagger}\,\partial_\mu U \rightarrow W_\mu$, and obtain the following HYM Lagrangian density
\begin{align}\label{eq:L_HYM}
    \mathcal{L}_{\mathrm{hym}} =& -\frac{1}{4}F^{a}_{\mu\nu}F^{\mu\nu}_a + \frac{g^2}{4}(v+\varphi)^2\,T^{a}_{b}\,W^{a}_{\mu}W^{\mu}_{b} - V_{\rm{ym}}\nonumber\\
    & + \frac{1}{2}\partial_{\mu}\varphi\,\partial^{\mu}\varphi - \lambda v^2 \varphi^2 - \lambda v \varphi^3 - \frac{1}{4}\lambda \varphi^4\,.
\end{align}
Here,
\begin{align}
    T^{ab} = \hat{n}^{T}\Bigl\{\frac{t^a}{2},\frac{t^b}{2}\Bigr\}\hat{n}
\end{align}
is a diagonal matrix having $(n-1)^2-1$ zero entries, $2(n-1)$ entries with the value $1/2$, and one entry with the value $(n-1)/n$. This implies that we end up with $(n-1)^2-1$ massless spin-1 dof, $2(n-1)$ massive spin-1 dof with mass $m = gv/2$, one massive spin-1 dof with mass $m' = gv\sqrt{(n-1)/2n}$, and the Higgs (spin-0) with mass $M_{\varphi} = v\sqrt{2\lambda}$.\\

For our purposes, we require the mass hierarchy $m,m' \ll M_{\varphi}$ (or equivalently $g \ll \sqrt{\lambda}$). Notice that this assumption is valid since it is consistent with quantum theory. The gauge coupling $g$, if small, tends to remain small under radiative/quantum corrections. This is because such correction terms are proportional to the bare value of $g$ itself. On the other hand, even if the vev $v$ (and therefore $M_{\varphi}$) was zero, the quartic self-coupling of the Higgs $\lambda$ generates a mass term for it, driving it away from zero. Another, although equivalent, argument in favor of this assumption comes from considering the behavior of global symmetries. While keeping  $M_\varphi = \sqrt{2\lambda}\,v$ fixed, all of the massive $W_{\mu}$ decouple with the Higgs in the limit $g \rightarrow 0$ (giving $m \rightarrow 0$), and we recover a global SU($n$) (note that there \textit{isn't} a global SU($n$) in the Higgs phase~\cite{Hertzberg:2018kyi,Hertzberg:2019ffc}). With this mass hierarchy then, we can integrate out the Higgs field to get an effective theory at smaller energy scales.\\

Upon integrating out the Higgs field and denoting $(m^2)^{b}_{a}$ as the mass squared matrix (which is a diagonal matrix with entries $\sim g^2v^2$), we get the following tree level Lagrangian density for the massive Yang-Mills sector
\begin{align}\label{eq:L_eff_relativistic}
    \mathcal{L}' &= -\frac{1}{4}F^2 + \frac{1}{2}(m^2)^{b}_{a}W^{a}_{\mu}W^{\mu}_{b} - V_{\rm{ym}} - V_{h}\,.
\end{align}
Here the Higgs induced potential is
\begin{align}\label{eq:Vh}
    V_{h}  &= - \frac{g^2}{2M_{\varphi}^2}\left(m^{b}_{a}\,W^{a}_{\mu}W^{\mu}_{b}\right)^2 + ...\,,
\end{align}
with ``$...$'' representing all the higher order derivative couplings (suppressed by powers of $M_{\varphi}$), which become irrelevant in as far as non-relativistic analysis is concerned. We shall discard them in this work. Notice the sign of the above interaction vertex. It dictates attractive self interaction between the vector fields (which was expected on account of it arising due to a spin-0 exchange). On the other hand, the quartic interaction vertex in $V_{\rm{ym}}$, which would be the relevant term in the non-relativistic regime, dictates a repulsive interaction. Due to the presence of both types of interactions along with gravity, we would have a diverse set of solitonic configurations (see Sec.~\ref{Sec:grav_bound} ahead).\\

In order for there to exist solitons in this setup, we require there to be no massless degrees of freedom in the spectrum of the theory. This is because any massive object, on account of the Yang-Mills interactions, would decay into such massless dof. In order to generate masses for the remaining massless spin-1 dof (for $n \geq 3$), we can `Higgs' them further. It is easy to see that in general for SU($n$), we would need $n^2-1$ real scalars. With different vevs, we can have a diverse set of mass values for the spin-1 dof, leading to interesting phenomenology. In this paper we shall focus on the simplest case of SU($2$), for which the generators $t^a$ are the three Pauli matrices $\tau^a$, and with the Levi-Civita symbols as structure constants
\begin{align}\label{eq:struc.const.SU2}
    f^{abc} = \varepsilon^{abc}\,.
\end{align}
The mass matrix becomes proportional to unity, i.e. identical masses for all the three spin-1 dof
\begin{align}
    (m^2)^{a}_{b} = \frac{g^2v^2}{2}T^{a}_{b} = \frac{g^2v^2}{4}\delta^{a}_{b} \equiv m^2\delta^{a}_{b}\,.
\end{align}
Extension to $n \geq 3$, including Higgs in the adjoint representation, is left for future work.\\

We end this section by commenting on the ratio of $m/M_{\varphi} = g/(2\sqrt{2\lambda})$. For the presentation in this paper, we shall assume their benchmark values to be $m = 10$ eV, and $M_{\varphi} = $ MeV. With $\lambda = \mathcal{O}(1)$,\footnote{Requiring perturbative Unitarity to hold at all energy scales accessible within the theory, we get an upper bound $\lambda(E) \leq 4\pi/3$. This is obtained by considering the $\ell = 0$ partial wave scattering in the $\phi + \bar{\phi} \rightarrow \phi + \bar{\phi}$ process at high energies.} we can have $g = \mathcal{O}(10^{-5})$ for these values of $m$ and $M_{\varphi}$, consistent with Bullet cluster observations (see Sec.~\ref{sec:g_bound} ahead for a bound on $g$ due to Bullet cluster observations). Although much smaller values of $g$ are allowed in principle (making the ratio $m/M_{\varphi}$ even smaller), one can wonder how could $g$ be so small if some unification is to take place with the known interactions of the SM at high energy scales. In this paper we remain agnostic towards such wonderment, and allow for all possible values of the gauge coupling (allowed by the Bullet cluster constraints). We shall see that even for such large values of the mass $m$, we can admit long lived Yang-Mills stars with varying sizes, ranging from macroscopic to even astrophysical (both in radii and masses). Such objects can carry very large amounts of spin and/or iso-spin.

\section{non-relativistic limit}
\label{Sec:NR_effective}

During its cosmological evolution, once the ambient energy scale in the dark sector becomes much smaller than the mass $m$, HYM goes through a transition where it becomes non-relativistic (with a conserved particle number due to suppressed number changing processes). For our purposes of studying non-topological soliton solutions, it is enough to work with Minkowski background. Extracting the Compton oscillations $e^{\pm imt}$, that is with the ansatz
\begin{align}\label{eq:W_decompose}
    {\bm W}({\bf x},t) = \frac{1}{\sqrt{2m}}\left[e^{-imt}\,\bPsi({\bf x},t) + h.c.\right],
\end{align}
and $W_{0}({\bf x},t) = \left[e^{-imt}\,\psi_0({\bf x},t) + h.c.\right]/\sqrt{2m}$ in~\eqref{eq:L_eff_relativistic}; where ${\bm W} = \tau_a{\bm W}^a/2$ giving ${\bPsi} = \tau_a{\bPsi}^a/2$; and keeping only the slowly oscillating piece in $\bPsi^a$ (slower than $m^{-1}$), one gets the following non-relativistic Lagrangian density\footnote{The second time derivative term and terms having $e^{\pm i nmt}$ oscillatory pieces are dropped. The constraint $\psi^a_0 = i(\partial_i\psi^a_i)/m$, obtained by varying the action with respect to $\psi^{a\,\ast}_0$, has been used to eliminate $\psi^a_0$ (and its conjugate $\psi^{a\,\ast}_0$). Dropping such oscillatory terms is equivalent to saying that number changing processes are suppressed. Equivalently, typical momenta/energy scales are much smaller than the mass, prohibiting creation/annihilation of a quanta of mass $m$.}
\begin{align}\label{eq:NR_action}
    \mathcal{L}_{\rm{nr}} &= \Re\bigl[i\,\bPsi^{a\dagger}\cdot\partial_t{\bPsi_a}\bigr] - \frac{1}{2m}\nabla_i\bPsi^{a\dagger}\cdot\nabla_i\bPsi_{a} + \mpl^2\,\Phi\nabla^2\Phi\nonumber\\
    &- m\,\Phi\,\bPsi^{a\dagger}\cdot\bPsi_a - V_{\rm ym,nr} - V_{\rm h,nr}\,.
\end{align}
Here $\mpl = (8\pi G)^{-1/2}$ is the reduced Planck mass and $\Phi = h_{00}/2$ is the Newtonian potential (with $h_{\mu\nu}$ being the metric perturbations around Minkowski $g_{\mu\nu} = \eta_{\mu\nu} + h_{\mu\nu}$\footnote{See~\cite{Adshead:2021kvl,Jain:2021pnk} for a detailed derivation of the Newtonian limit of linearized gravity for the case of a spin-$1$ field. Here we have the exact same situation, with the only difference of having three spin-$1$ fields with self interactions.}). The interaction potentials are
\begin{align}\label{eq:NR_potentials}
    V_{\rm{ym,nr}} =& \,\frac{g^2}{8m^2}\varepsilon^{abc}\varepsilon_{ade}\Bigl[(\bPsi^{d\dagger}\cdot\bPsi^{b})(\bPsi^{e\dagger}\cdot\bPsi^{c})\nonumber\\
    &+ (\bPsi^{d}\cdot\bPsi^{b})(\bPsi^{e\dagger}\cdot\bPsi^{c\dagger}) + (\bPsi^{d\dagger}\cdot\bPsi^{b})(\bPsi^{e}\cdot\bPsi^{c\dagger})\Bigr]\nonumber\\
    V_{\rm{h,nr}} =& - \frac{g^2}{4M_{\varphi}^2}\Big[(\bPsi^a\cdot\bPsi^a)(\bPsi^{b\dagger}\cdot\bPsi^{b\dagger})\nonumber\\
    &+ 2\,(\bPsi^a\cdot\bPsi^{a\dagger})(\bPsi^{b}\cdot\bPsi^{b\dagger})\Bigr]\,,
\end{align}
where the cubic interaction term in the Yang-Mills potential, on account of being suppressed by additional factors of $|\nabla|/m$, drops out in the non-relativistic limit. This gives the following non-Abelian Schr\"{o}dinger-Poisson system of equations
\begin{align}\label{eq:masterSP1}
    i\partial_t\bPsi^a = &-\frac{1}{2m}\nabla^2\bPsi^a + m\,\Phi\,\bPsi^a + \frac{\partial V_{\rm{ym,nr}}}{\partial\bPsi^{a\dagger}} + \frac{\partial V_{\rm{h,nr}}}{\partial\bPsi^{a\dagger}}\nonumber\\
    \nabla^2\Phi = &\frac{m}{2\mpl^2}\bPsi^{a\dagger}\cdot\bPsi^a\,.
\end{align}
We note that apart from the presence of self interactions induced due to heavy Higgs ($V_{\rm h, nr}$) and Yang-Mills structure ($V_{\rm ym, nr}$), the Lagrangian density~\eqref{eq:NR_action} (and hence the Schrodinger Poisson system~\eqref{eq:masterSP1}) is the same as that of nine non-relativistic equal mass fields, interacting with Newtonian gravity through the mass density $m\,\bPsi^{a\dagger}\cdot\bPsi_a$. This is as expected. The self interactions add another layer to the dynamics, which is specific to the UV structure of the theory. Lastly, extension to the FLRW universe (with $a$ and $H=\dot{a}/a$ being the scale factor and the Hubble parameter), can be made by replacing $\nabla\rightarrow \nabla/a$ and $\partial/\partial t\rightarrow \partial/\partial t+3H/2$.\footnote{The factor of $3/2$ reflects the fact that the non-relativistic field $\bPsi$ red-shifts as matter, i.e. it red-shifts as $a^{-3/2}$.}

\subsection{Conserved quantities}

With the NR theory~\eqref{eq:NR_action} at hand, we now discuss its various symmetries and the associated conserved quantities. To begin, the system has a global $U(1)$ symmetry ($\bPsi\rightarrow e^{i\alpha}\bPsi$) leading to a conserved particle number
\begin{align}\label{eq:total_number}
    N = \int \mathrm{d}^3 x \,\bPsi^{a\dagger}\cdot\bPsi^a\,.
\end{align}
Note that the presence of non-gravitational interaction potentials~\eqref{eq:NR_potentials} prohibit separate global U($1$)s, both across different vector/color fields, and across different components (or equivalently spin multiplicity fields~\cite{Jain:2021pnk}) of a given vector field. This is (and as expected) similar to the case of a single vector field with non-gravitational self interactions~\cite{Zhang:2021xxa}. Next, there are various conserved quantities associated with Galilean invariance of the NR system. The energy (charge associated with time translation invariance) evaluates to be
\begin{align}\label{eq:Energy}
    H =& \int\mathrm{d}^3x\Biggl[\frac{1}{2m}\nabla_i\bPsi^a\cdot\nabla_i\bPsi^{a\,\dagger} + \frac{1}{2}m\,\Phi\,\bPsi^{a\,\dagger}\cdot\bPsi^a\nonumber\\
    &\qquad\qquad + V_{\rm{ym,nr}} + V_{\rm{h,nr}}\Biggr]\,,
\end{align}
where $\Phi$ is the solution of the Poisson equation in~\eqref{eq:masterSP1}:
\begin{align}
    \Phi({\bf x},t) = -\frac{m}{2\mpl^2}\int\frac{\mathrm{d}^3y}{4\pi|{\bf x}-{\bf y}|}\,\bPsi^{a\dagger}({\bf y},t)\cdot\bPsi^a ({\bf y},t)\,.\nonumber
\end{align}
The first, second, and the last two terms in~\eqref{eq:Energy} are identified as $H_{\mathrm{press}}$, $H_{\mathrm{grav}}$, and $H_{\mathrm{self}}$ for the energy due to gradient pressure, gravity, and self interactions respectively. The next set of transformations in the Galileo group are spatial transformations. Invariance under spatial translation gives rise to linear momentum conservation. More importantly on the other hand, invariance under spatial rotation $\bPsi^a({\bf x}) \rightarrow \bR\cdot\bPsi^a(\bR^{-1}\cdot{\bf x})$ (where $\bR$ is a rotation matrix), gives rise to spin and orbital angular momentum
\begin{align}\label{eq:spin__ang_mom}
    {\bS}_{\mathrm{tot}} &= i\int\mathrm{d}^3x\,\bPsi^a\times\bPsi^{a\,\dagger},\\
    {\bf L}_{\mathrm{tot}} &= \int\mathrm{d}^3x\,\Re\left[i\,\left(\bPsi^{a\,\dagger}\cdot\nabla\bPsi^a\right)\times {\bf x}\right]\,.
\end{align}
Of particular interest to us is the spin angular momentum. Notice that it is the \textit{total} spin (and total orbital angular momentum) of all the three SU($2$) fields that is conserved, but not individually. Last but not the least, is the custodial global SU($2$) symmetry. The transformation $\bPsi \rightarrow U^{\dagger}\,\bPsi\,U$ where $U = \rm{exp}(i\theta^a \tau_a)$ is a global SU($2$) matrix, leaves the action invariant. This gives rise to the conserved iso-spin
\begin{align}
    I_a = i\,\varepsilon_{abc}\int\mathrm{d}^3x\,\bPsi^b\cdot\bPsi^{c\,\dagger}\,.
\end{align}
Using a vector like notation, we shall denote this iso-spin as a $1 \times 3$ tuple (suitable for our SU($2$) case), with the three entries corresponding to the three components:
\begin{align}\label{eq:I_convention}
    I \equiv (I_1,I_2,I_3)\,.
\end{align}

\subsubsection{Accidental degeneracy}
\label{subsubsec:accid_sym}
Apart from the symmetries discussed above, there is an additional accidental symmetry that arises in the case of SU($2$), due to it being isomorphic to SO($3$) (and both the spatial and internal iso-indices taking values from $1$ to $3$). Notice from above that the spin and iso-spin are indistinguishable. So for any field configuration carrying some amount of total spin and iso-spin (even zero), there could potentially exist other configurations where the two quantities are swapped. In table~\ref{tab:SU(2)_solitons} we show two degenerate (in energy) pairs of soliton solutions due to this indistinguishability.

\subsection{Bound on self-interactions}\label{sec:g_bound}

If the HYM sector is to constitute the observed dark matter, we must make sure that it satisfies observational constraints on the dark matter self interactions. In this sub-section, we provide an estimate for the bound on the gauge coupling $g$, owing to Bullet cluster observations. Given the non-relativistic Yang-Mills interaction potential $V_{\mathrm{ym},\mathrm{nr}}$~\eqref{eq:NR_potentials}, scattering cross section in a non-relativistic elastic collision is\footnote{In general, scattering cross section in a $2 \rightarrow 2$ process is given as~\cite{Peskin:1995ev} $d\sigma/dt = |\mathcal{M}|^2/(64\pi s |{\bf p}_{i,cm}|)$ where $s$ and $t$ are Mandelstam variables, ${\bf p}_{i,cm}$ is the momentum of an incoming particle, and $|\mathcal{M}|$ is the matrix element.}
\begin{align}\label{eq:cross_section}
    \sigma = \frac{g^4|\mathcal{A}|^2}{64\pi m^2}\,,
\end{align}
where $|\mathcal{A}| \sim \mathcal{O}(1)$ is the dimensionless matrix element (without the $g^{2}$ vertex factor). We can estimate $|\mathcal{A}|$ under some assumptions: Consider free plane waves
\begin{align}
    \bPsi^{a} \propto {\bm \beta}^a\,e^{i\left({\bf k}_a\cdot{\bf x}-\frac{k^2_a}{2m}t\right)}\,,
\end{align}
where $|{\bm \beta}^a|$ gives the (dimensionless) amplitude of waves. The matrix element for this Yang-Mills potential~\eqref{eq:NR_potentials} is
\begin{align}\label{eq:A_matrixelement}
    |\mathcal{A}| =& \Bigl[\varepsilon^{abc}\varepsilon_{ade}\delta_{ij}\delta_{kl} + \varepsilon^{abd}\varepsilon_{ace}\delta_{ik}\delta_{jl}\nonumber\\
    &\qquad\qquad + \varepsilon^{abe}\varepsilon_{adc}\delta_{ij}\delta_{kl}\Bigr]\beta^{d\ast}_i \beta^{e\ast}_k \beta^{b}_j \beta^{c}_{l}\,.
\end{align}
Assuming equipartition among all dof (which may be a reasonable assumption for the state of dark matter throughout most of the halos in the Bullet cluster), ${\bm \beta}^a$ can be taken to be random complex numbers (each of the phases distributed uniformly in $[0,2\pi)$) with magnitude $1/3$. That is,
\begin{align}
    \langle \beta^{d\ast}_i \beta^{e\ast}_k \beta^{b}_j \beta^{c}_{l}\rangle =& \frac{1}{3^4}\Bigl[\delta_{ij}\delta_{kl}\delta^{db}\delta^{ec} + \delta_{il}\delta_{jk}\delta^{dc}\delta^{be}\nonumber\\
    &\qquad\qquad\qquad + \delta_{ijkl}\delta^{bcde}\Bigr]\,.
\end{align}
Using this in~\eqref{eq:A_matrixelement} gives $\langle|\mathcal{A}|\rangle = 8/9$ for our SU($2$) case.\\

Finally, Bullet cluster observations suggest $\sigma/m \lesssim \eta$ cm$^2/$g where $\eta$ is an order unity parameter depending upon the analysis~\cite{Markevitch:2003at,Harvey:2015hha}. This translates to the following upper bound on the self-coupling
\begin{align}\label{eq:g_bound}
    g \lesssim 3.3\times 10^{-5}\,\eta^{1/4}\,\left(\frac{m}{10\,\mathrm{eV}}\right)^{3/4} \equiv g_{max}\,.
\end{align}
Throughout the rest of this work, we will set $\eta = 1$ for concreteness. For any other value, $g_{\rm max}$ can be trivially rescaled using its definition above.\\

\section{Types of soliton solutions: Yang-Mills stars}
\label{Sec:grav_bound}

Owing to the wave-like behavior of the field~\eqref{eq:masterSP1} together with gravitational and self-interactions, perturbations (on top the background field) in the field can come together and form coherently oscillating objects with large occupation numbers, called solitons.
In this section we discuss the various soliton solutions, `Yang-Mills stars', admitted in our HYM model. We look for field solutions that minimize the energy functional~\eqref{eq:Energy} at some fixed particle number~\eqref{eq:total_number}. That is, we extremize the quantity
\begin{align}\label{eq:scriptE}
    \mathcal{E} = H + \frac{\mu}{m} (M-M_s)
\end{align}
with respect to $\bPsi^a$, where $\mu > 0$ is the chemical potential (a Lagrange multiplier). This leads to the time independent Schr\"{o}dinger equation ($i\partial_t\bPsi^a \rightarrow -\mu\bPsi^a$ in~\eqref{eq:masterSP1}), dictating coherent field configurations. Focusing on spherically symmetric soliton profiles (not necessarily the field solutions), we therefore consider the following ansatz
\begin{align}\label{eq:master_ansatz}
    \bPsi = \bPsi^a({\bf r},t)\,\tau_a = e^{i\mu t}\,\psi(r)\,\alpha_{a}\,\bepsilon^{(\lambda_a)}_{\hat{n}_a}\,\tau_a\,.
\end{align}
Here, $\alpha_a$ are arbitrary (complex) coefficients, and the set $\{\hat{\bepsilon}^{(\lambda_a)}_{\hat{n}_a}\}$ are the polarization eigenvectors with the spin multiplicity value equal to $\lambda_a$ in the $\hat{n}_a$ direction. This corresponds to having a particular type of polarization (linear $\lambda_a = 0$, or circular $\lambda_a = \pm 1$) along some direction $\hat{n}_a$, for each of the different vector fields in the internal iso-spin space.\footnote{For details, see~\cite{Jain:2021pnk} for the case of a single vector field.} In general, $\{\hat{\bepsilon}^{(\lambda_a)}_{\hat{n}_a}\}$ form an orthonormal basis set in the following sense 
\begin{align}
    \bepsilon^{(\lambda_a)\dagger}_{\hat{n}_a}\cdot\bepsilon^{(\lambda'_a)}_{\hat{n}_a} = \delta^{\lambda_a\lambda'_a}\;;\;\sum_{\lambda_a}\Big[\bepsilon^{(\lambda_a)\dagger}_{\hat{n}_a}\,\bepsilon^{(\lambda_a)}_{\hat{n}_a}\Big]_{ij} = \delta_{ij}\,.
\end{align}
For $\hat{n} = \hat{z}$, we have $\bepsilon^{(0)}_{\hat{z}} = \hat{\bm z}$, $\bepsilon^{(\pm 1)}_{\hat{z}} = (\hat{\bm x} \pm i\hat{\bm y})/\sqrt{2}$. Without loss of generality, we set 
\begin{align}
    \sum_a|\alpha_a|^2 = 1
\end{align}
in order to have the total particle number independent of $\alpha_a$, i.e.
\begin{align}
    N = \int\mathrm{d}^3x \,\psi^2\,.
\end{align}
Using the ansatz~\eqref{eq:master_ansatz} in~\eqref{eq:masterSP1}, we get the following Schr\"{o}dinger-Poisson system 
\begin{align} \label{eq:soliton_master_eq}
     -\mu\psi\,\alpha_a\,\bepsilon^{(\lambda_a)}_{\hat{n}_a} =& \left(-\frac{1}{2m}\nabla^2\psi + m\,\Phi\,\psi\right)\alpha_a\,\bepsilon^{(\lambda_a)}_{\hat{n}_a}\nonumber\\
     &+ \frac{g^2}{4m^2}\left({\bm \kappa^a_{\rm{ym}}} - \frac{2m^2}{M_{\varphi}^2}{\bm \kappa}^a_{h}\right)\psi^3\nonumber\\
    \nabla^2\Phi =& \frac{m}{2\mpl^2}\psi^2\,,
\end{align}
with $\nabla^2f(r) = f''(r) + (2/r)f'(r)$ for a radially symmetric function $f(r)$, and
\begin{align}
    {\bm \kappa}^a_{\mathrm{ym}} =& \frac{1}{2}\,\varepsilon^{gbc}\varepsilon_{gde}\Bigl[2\,\delta^{da}\alpha_{b}\alpha^{\ast}_{e}\alpha_{c}(\bepsilon^{(\lambda_e)\dagger}_{\hat{n}_e}\cdot\bepsilon^{(\lambda_c)}_{\hat{n}_c})\bepsilon^{(\lambda_b)}_{\hat{n}_b}\nonumber\\
    &\qquad\qquad\quad + \delta^{ca}\alpha^{\ast}_{e}\alpha_{d}\alpha_{b}(\bepsilon^{(\lambda_d)}_{\hat{n}_d}\cdot\bepsilon^{(\lambda_b)}_{\hat{n}_b})\bepsilon^{(\lambda_e)\dagger}_{\hat{n}_e}\nonumber\\
    &\qquad\qquad\quad + \delta^{ea}\alpha^{\ast}_{c}\alpha_{d}\alpha_{b}(\bepsilon^{(\lambda_d)}_{\hat{n}_d}\cdot\bepsilon^{(\lambda_b)}_{\hat{n}_b})\bepsilon^{(\lambda_c)\dagger}_{\hat{n}_c}\nonumber\\
    &\qquad\qquad\quad +\delta^{da}\alpha^{\ast}_{c}\alpha_{e}\alpha_{b}(\bepsilon^{(\lambda_e)}_{\hat{n}_e}\cdot\bepsilon^{(\lambda_c)\dagger}_{\hat{n}_c})\bepsilon^{(\lambda_b)}_{\hat{n}_b}\nonumber\\
    & \qquad\qquad\quad + \delta^{ca}\alpha^{\ast}_{d}\alpha_{b}\alpha_{e}(\bepsilon^{(\lambda_d)\dagger}_{\hat{n}_d}\cdot\bepsilon^{(\lambda_b)}_{\hat{n}_b})\bepsilon^{(\lambda_e)}_{\hat{n}_e}\Bigr]\nonumber\\
    {\bm \kappa}^a_{h} =& \,\alpha_{b}\alpha_{b}\delta_{\lambda_{b},0}\,\alpha^{\ast}_{a}\bepsilon^{(\lambda_a)\dagger}_{\hat{n}_a} + 2\,\alpha_{a}\bepsilon^{(\lambda_a)}_{\hat{n}_a}\,.
\end{align}

\begin{figure*}[t]
\begin{center}
\includegraphics[width=0.9\textwidth]{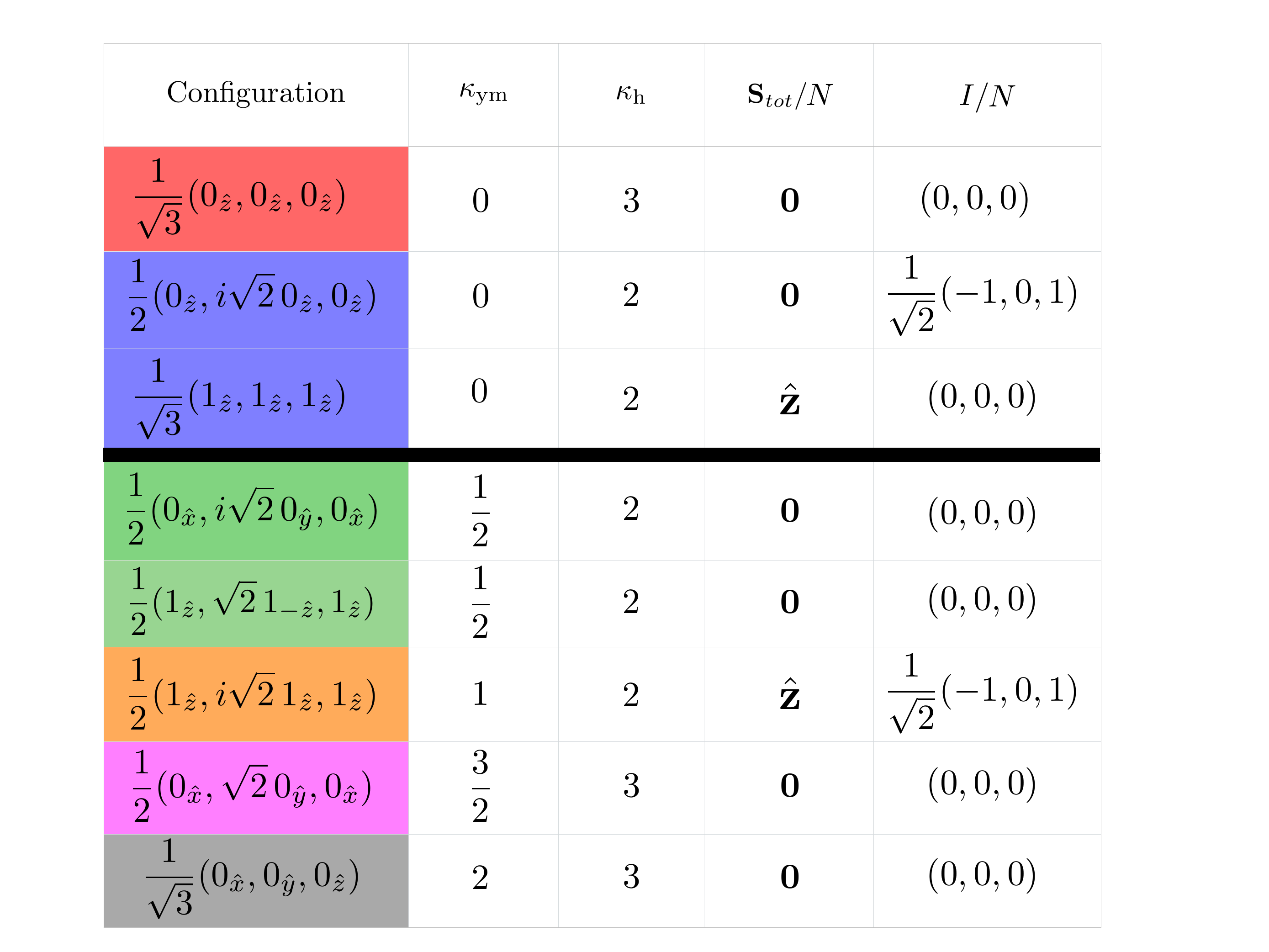}
\end{center}
\caption{A table of solitons in a Higgsed SU($2$) theory. As explained in the text, $1_{\hat{n}}$ and $0_{\hat{n}}$ represent circular polarization and linear polarization along the $\hat{n}$ direction respectively, while the three entries correspond to the three directions in the internal iso-spin space. The second and third columns give the coefficient $\kappa_{\rm{ym}}$ and $\kappa_{\rm{h}}$ of the Yang-Mills and Higgs induced term in~\eqref{eq:kappa_eff} respectively. The last two columns give the (conserved) spin angular momentum and iso-spin, for each configuration. Configurations with the same value of $\kappa_{\mathrm{ym}}$ and $\kappa_{\mathrm{h}}$ are degenerate (reflected by the same color representation). The upper set corresponds to solitons with attractive self-interactions within them, while the lower set corresponds to ones having repulsive self-interactions The color coding is the same as in fig.~\ref{fig:HandRvsN}. Since any soliton solution spontaneously breaks the global SO($3$) and SU($2$) symmetry of the effective non-relativistic theory~\eqref{eq:NR_action}, an infinite number of solitons can be trivially obtained by performing rotations in physical and/or internal iso-spin space. Also note the two degenerate pairs of soliton solutions (blue and green color), arising in our special case of SU($2$). See~\ref{subsubsec:accid_sym} for a discussion. Note that the first and the third rows are basically single vector solitons in a U($1$) Abelian Higgs model (see Sec.~\ref{sec:Abelian.special.case} for a discussion). This can be seen by performing an iso-rotation (rotation in the internal iso-spin space) to have all but one entry non-zero. 
}
\label{tab:SU(2)_solitons}
\end{figure*}

In order for our ansatz to be consistent, we require \textit{both} ${\bm \kappa}^a_{\rm{ym}}$ and ${\bm \kappa}^a_{\rm{h}}$ to be proportional to $\alpha_a\bepsilon^{(\lambda_a)}_{\hat{n}_a}$:\footnote{Note that in the case when ${\bm \kappa}^a_{\rm{ym}} \neq 0$, requiring ${\bm \kappa}^a_{\rm{h}} = \kappa_{\rm{h}}\,\alpha_a\bepsilon^{\lambda_a}_{\hat{n}_a}$ may not be necessary since the contribution from $V_{\varphi}$ is suppressed by $(m/M_{\varphi})^2 \ll 1$. This opens up more possibilities for solitonic solutions, which may or may not be stable depending upon their specific configuration. We leave a detailed analysis of such configurations for future work.} 
\begin{align}\label{eq:kappa_req}
    {\bm \kappa}^a_{\rm{ym}} = \kappa_{\rm{ym}}\,\alpha_a\bepsilon^{(\lambda_a)}_{\hat{n}_a}\nonumber\\
    {\bm \kappa}^a_{\rm{h}} = \kappa_{\rm{h}}\,\alpha_a\bepsilon^{(\lambda_a)}_{\hat{n}_a}\,.
\end{align}
This places severe constraints on $\alpha_a$ and $(\lambda_a,\hat{n}_a)$, and therefore on the type of solitons we can have. With~\eqref{eq:kappa_req}, the Schr\"{o}dinger-Poisson system for the radially symmetric profile $\psi(r)$ is
\begin{align}\label{eq:profile_master_eq}
    -\mu\psi = & -\frac{1}{2m}\nabla^2\psi + m\,\Phi\,\psi + \frac{g^2}{4m^2}\,\kappa_{\mathrm{eff}}\,\psi^3\nonumber\\
    \nabla^2\Phi = &\frac{m}{2\mpl^2}\psi^2\,,
\end{align}
along with the following total spin and iso-spin
\begin{align}\label{eq:spiniso-spin_sol}
    {\bS}_{\mathrm{tot}} &= i\,|\alpha_a|^2\left(\bepsilon^{(\lambda_a)}_{\hat{n}_a}\times\bepsilon^{(\lambda_a)\dagger}_{\hat{n}_a}\right)\,N\\
    I_a &= i\,\varepsilon_{abc}\,\alpha_{b}\alpha^{\ast}_{c}\left(\bepsilon^{(\lambda_b)}_{\hat{n}_b}\cdot\bepsilon^{(\lambda_c)\dagger}_{\hat{n}_c}\right)N\,.
\end{align}
The energy is
\begin{align}\label{eq:energy_sol}
    H &= \int\mathrm{d}^3x\Biggl[\frac{1}{2m}\left(\nabla\psi\right)^2 + \frac{1}{2}m\Phi\psi^2 + \frac{1}{8}\frac{g^2\kappa_{\mathrm{eff}}}{m^2}\psi^4\Biggr],
\end{align}
where $\Phi$ is given by the solution of the Poisson equation in~\eqref{eq:profile_master_eq}:
\begin{align}
    \Phi({\bf x}) = -\frac{m}{2\mpl^2}\int\frac{\mathrm{d}^3y}{4\pi|{\bf x}-{\bf y}|}\,\psi^2({\bf y})\,.\nonumber
\end{align}
We have defined an effective $\kappa$
\begin{align}\label{eq:kappa_eff}
    \kappa_{\mathrm{eff}} \equiv \kappa_{\rm{ym}} - \left(\frac{2m^2}{M_{\varphi}^2}\right)\,\kappa_{\rm{h}}\,,
\end{align}
dictating what type of interactions (attractive or repulsive) reside within a soliton configuration: Depending upon the specific choices for $\{\alpha_{a},\lambda_{a},\hat{n}_{a}\}$ such that \eqref{eq:kappa_req} is satisfied, we can have both $\kappa_{\mathrm{ym}} \neq 0$ and $\kappa_{\mathrm{ym}} = 0$, while $\kappa_{h} \neq 0$ always. The Higgs induces an attractive self interaction, while the inherent Yang-Mills interaction is repulsive (with $m^2/M^2_{\varphi} \ll 1$). Hence, we have both classes of solitons studied in the literature (i.e. having attractive or repulsive interactions alongside gravity), albeit mostly in the context of scalar fields~\cite{Schive:2014dra, Schiappacasse:2017ham,Hertzberg:2018lmt,Croon:2018ybs,Amin:2019ums,Arvanitaki:2019rax,Hertzberg:2020xdn,Salehian:2021khb}. (See~\cite{Aoki:2017ixz,Adshead:2021kvl,Jain:2021pnk,Zhang:2021xxa} for solitons/oscillons in single spin-1 and higher spin fields without repulsive self interactions). Owing to the important distinction of the spin-1 nature of the vector fields and the Yang-Mills structure, the HYM solitons can carry large intrinsic spin and/or iso-spin.\\

The Schr\"{o}dinger Poisson system for $\psi(r)$~\eqref{eq:profile_master_eq} is the same as that obtained for a single scalar field. Reproducing some of main results in the literature~\cite{Chavanis:2011zi,Chavanis:2011zm,Schiappacasse:2017ham,Hertzberg:2020xdn}, we will discuss the various possible soliton solutions (with a radially symmetric field profile and hence zero orbital angular momentum) in the  SU($2$) HYM theory. Depending upon the value of $\kappa_{\mathrm{eff}}$, each class (attractive if $\kappa_{\mathrm{eff}} < 0$, or repulsive if $\kappa_{\mathrm{eff}} > 0$) admits various soliton solutions with different energies. In Table~\ref{tab:SU(2)_solitons} we enumerate the various types of possible solitons in SU($2$). We have represented the different solitons using the polarization basis~\eqref{eq:master_ansatz}, with the three entries in the 3-tuple corresponding to the three vector fields. In general, the entry $1_{\hat{n}}$ in a slot corresponds to that particular field being circularly polarized along the $\hat{n}$ direction (i.e. having spin multiplicity $\lambda = 1$ in that direction). Similarly $0_{\hat{n}}$ represents linear polarization ($\lambda = 0$) along $\hat{n}$ direction.\\

\begin{figure*}[t]
\begin{center}
\includegraphics[width=1\textwidth]{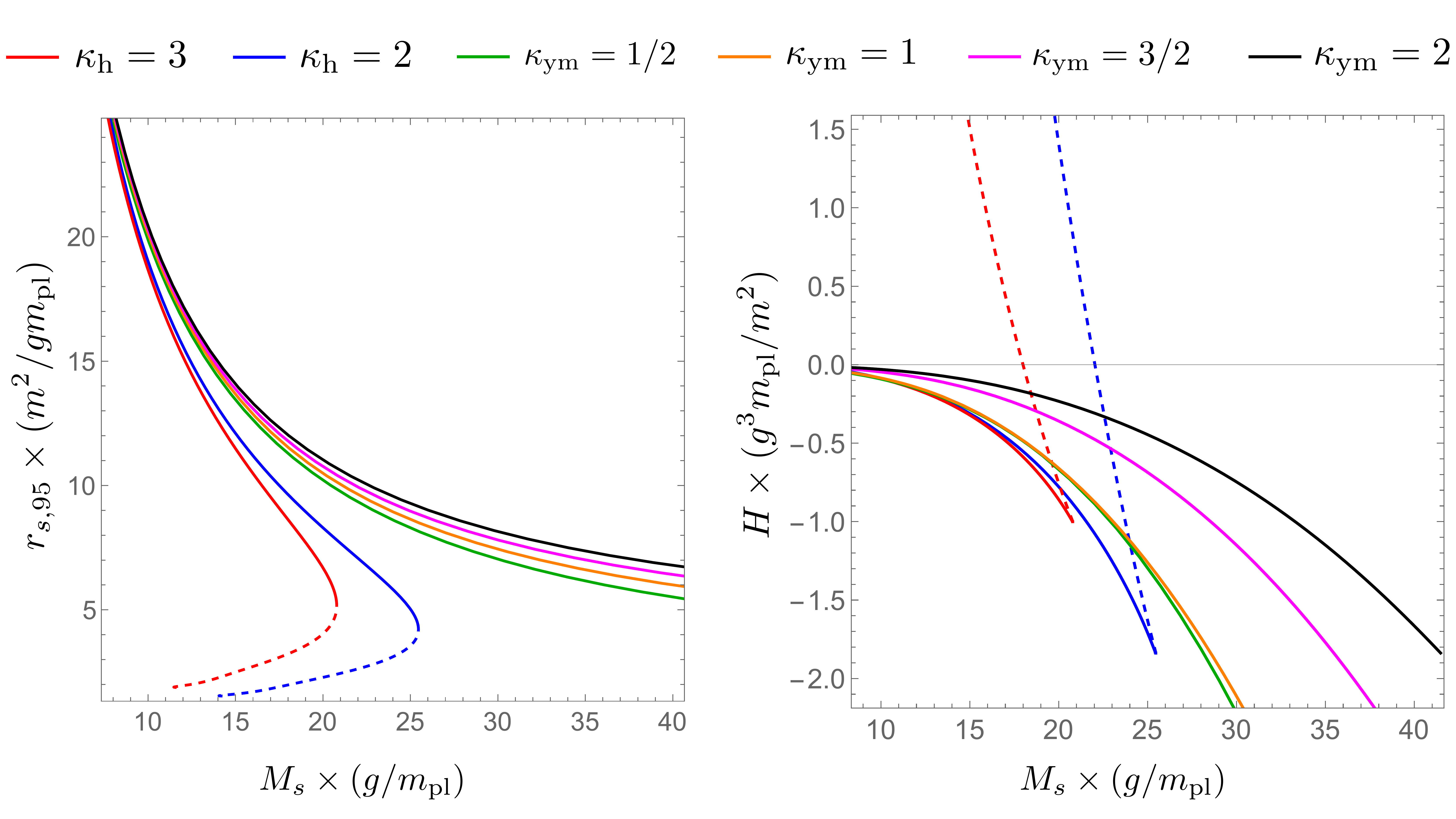}
\end{center}
\caption{Radius $r_{s,95}$ vs mass $M_s$ curves (left panel), and energy $H$ vs $M_s$ curves (right panel), for the six different types of solitons in the SU(2) HYM, discussed in this paper. Color coding is the same as in Table~\ref{tab:SU(2)_solitons}. Starting on any curve from the upper left corner of the left panel, or from the middle left of the right panel, and traversing towards the right along it, the chemical potential $\mu$ increases. Red and blue correspond to soliton families with attractive self interactions within them (with the replacement $g \rightarrow g\sqrt{2}m/M_{\varphi}$ in the scaling of axes). The dashed part of the curves correspond to unstable configurations. On the other hand the orange, green, magenta, and black curves correspond to soliton families with repulsive self-interactions within them. Both the blue and orange families carry large amounts of spin and/or iso-spin. Extending towards the right (higher values of $M_s$) for $\kappa_{\mathrm{ym}} \neq 0$ solitons, connects to fig.~\ref{fig:HandRvsN2}.}
\label{fig:HandRvsN}
\end{figure*}

To get the scaling relations between the masses and radii of solitons in both the classes, we will use $\nabla^{2} \sim 1/r_s^2$ and $m\psi^2 \sim M_s/r_s^3$, where $r_s$ and $M_s$ are the radius and mass of the soliton(s) respectively. Using this, the various terms in the total energy of the soliton $H = H_{\mathrm{press}} + H_{\mathrm{grav}} + H_{\mathrm{self}}$ (c.f.~\eqref{eq:energy_sol}) scale like
\begin{align}
    \label{eq:energies_gp}
    H_{\mathrm{press}} &\sim \frac{M_s}{2m^2r_s^2}\,,\\
    \label{eq:energies_grav}
    H_{\mathrm{grav}} &\sim -\frac{M_s^2}{4\mpl^2r_s}\,,\\
    \label{eq:energies_self}
    H_{\mathrm{self}} &\sim \frac{g^2\kappa_{\mathrm{eff}}}{8m^4}\frac{M_s^2}{r_s^3}\,,
\end{align}
with the three being due to gradient pressure, gravity, and self interactions respectively. Since the full family of soliton solutions (labelled by different values of the chemical potential $\mu$) correspond to the extremum of $\mathcal{E}$~\eqref{eq:scriptE}, we also have
\begin{align}\label{eq:delHdelN_mu}
    \frac{\delta H}{\delta M_s} = -\frac{\mu}{m}\,.
\end{align}
This would fetch the relationship between the various soliton masses (or equivalently radii) and the associated chemical potential, and can be used to guess the goodness of non-relativistic approximation. The smaller the ratio $\mu/m$, the slower the Schr\"{o}dinger fields' oscillations as compared to Compton time $m^{-1}$, hence better the approximation. It is to be noted that $\delta$ in the above expression represents variation with respect to the mass $M_s$ (and not the field $\psi$). For solitons with attractive self interactions, $H(M_s)$ cannot be defined uniquely, for in this case there is a maximum value of $M_s$ that can be achieved. This corresponds to the transition point after which solitons start to become unstable (as $\mu$ increases), and is reflected as a kink in the $H (M_s)$ curve. (See the behavior of red and blue curves in the right panel of Fig.~\ref{fig:HandRvsN} ahead.) However on either side of this transition point, $H(M_s)$ is single valued and~\eqref{eq:delHdelN_mu} is well defined.\\

\begin{figure*}[t]
\begin{center}
\includegraphics[width=1\textwidth]{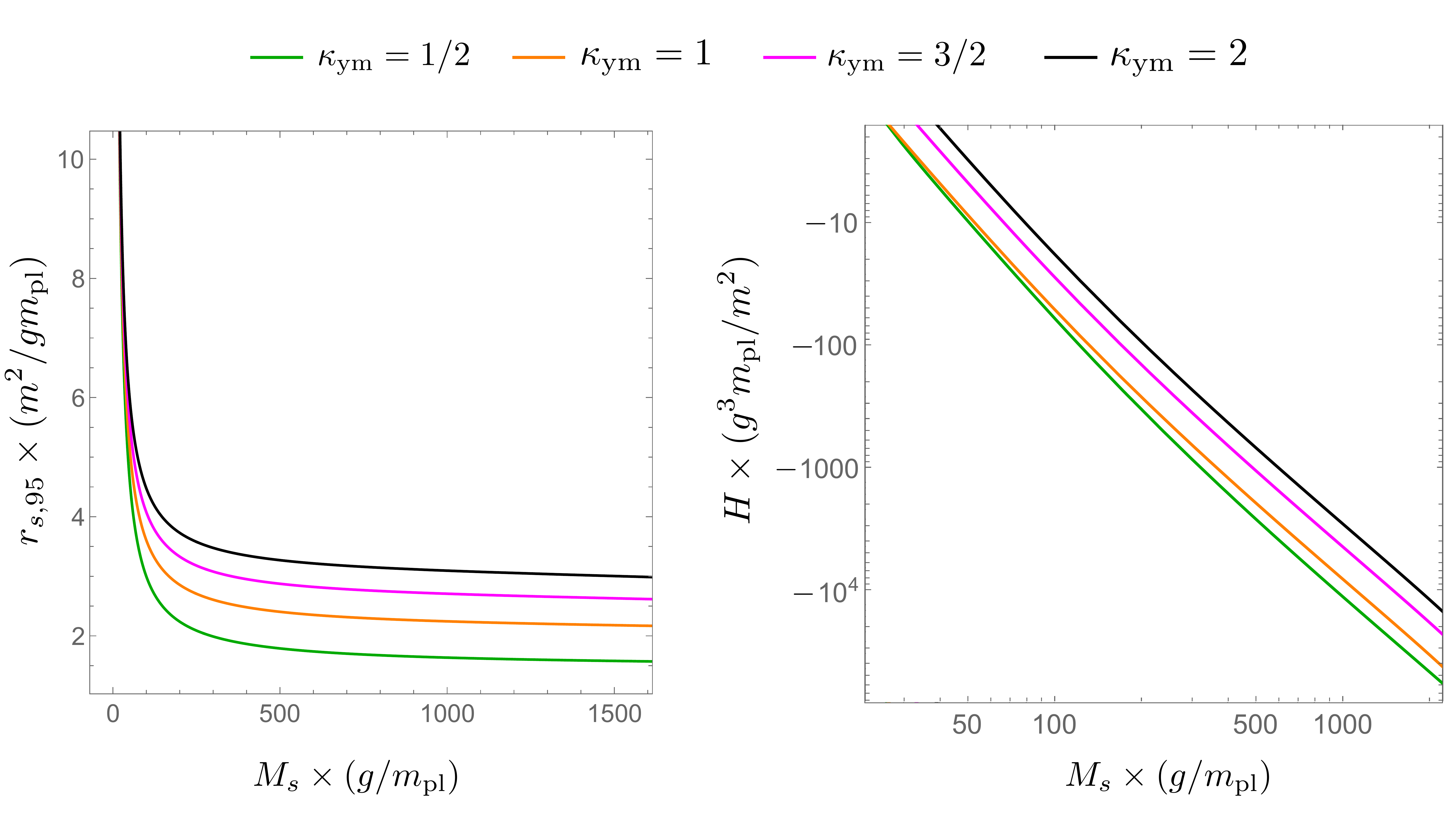}
\caption{Radius $r_{s,95}$ vs mass $M_s$ curves (left panel), and energy $H$ vs $M_s$ curves (right panel), for solitons with repulsive self interactions. The left panel shows that the radius of the soliton becomes almost independent of its mass at large values, giving rise to the Chandrasekhar regime~\eqref{eq:Ms_upper}. Since in this regime $H \propto M_s^2$ (which is also seen towards the right hand side of the right panel), we can extrapolate any curve on the right panel using~\eqref{eq:HvsM_extrapolate}, to get an estimate for the energies of more massive Chandrasekhar solitons (c.f.~\eqref{eq:HvsM_extrapolate}).}
\label{fig:HandRvsN2}
\end{center}
\end{figure*}

To understand the two different classes of soliton solutions, note that there are basically three regimes. First is the dilute soliton regime where the gradient pressure is balanced by gravity while self interactions are only sub-dominant. In this regime, both of the classes approach each other (the upper left corner of the left panel in fig.~\ref{fig:HandRvsN}). Second regime corresponds to having attractive self-interactions which balance out the gradient pressure (with gravity being sub-dominant). This lies towards the lower left corner of the left panel in fig.~\ref{fig:HandRvsN}. Lastly, the third regime is where self-interactions are repulsive and balance out gravity (with gradient pressure being sub-dominant). This is the lower right part of the left panel of fig.~\ref{fig:HandRvsN}, or equivalently the region where the curves flatten out in the left panel of fig.~\ref{fig:HandRvsN2}.\\

In the following we discuss these three difference regimes in detail. Moreover, upon including quantum mechanical effects, it has been suggested that there are small scale perturbations (with wave-numbers comparable to the mass $m$) within solitons, that can lead to number changing processes due to non-gravitational self interactions~\cite{Hertzberg:2010yz}. This can lead to perturbative decay of solitons. More dramatically, perturbations may also grow exponentially for dense enough solitons, owing to the phenomenon of parametric resonance~\cite{Hertzberg:2018zte,Hertzberg:2014jza,Hertzberg:2020xdn}. We discuss the lifetimes of our solitons against these effects later in subsections~\ref{sec:pert_decay} and~\ref{sec:resonance_decay}.\\

\begin{figure}[!htb]
\begin{center}
\includegraphics[width=0.48\textwidth]{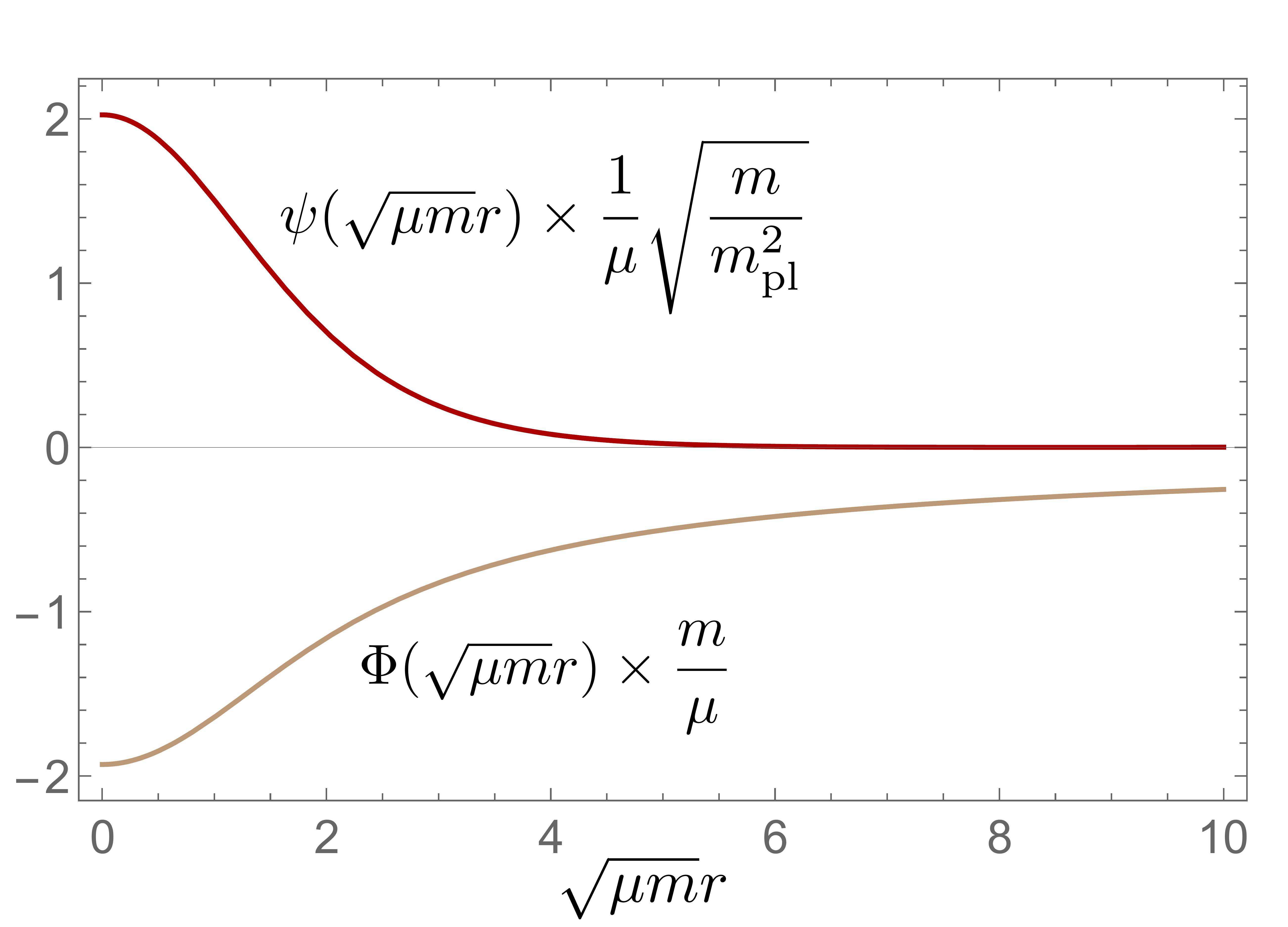}
\end{center}
\caption{Re-scaled scalar field profiles $\psi(r)$ and $\Phi(r)$ for a soliton 
held together by a balance between gravity and gradient pressure to leading order (lying in regime I). The numerical value of the combination $\mu g^2\kappa_{\mathrm{eff}}\mpl^2/(4m^3) = 0.001$ for these profiles, which do not exhibit any visible differences for $-0.001$. This soliton resides towards the upper left corner of the left panel of fig.~\ref{fig:HandRvsN}. Depending upon the specific value of $\kappa_{\rm eff}$ (for a given set of $\mu$, $g$ and $m$), it contains intrinsic spin and/or iso-spin (see Table~\ref{tab:SU(2)_solitons}).} \label{fig:field_profiles1}
\end{figure}

\noindent\textbf{Regime I: Gravity balancing gradient pressure.} With either attractive or repulsive self interactions, this regime corresponds to smaller values of $\mu$ (for a fixed value of $g$, $m$, $M$, and $\kappa_{\mathrm{eff}}$). Balancing the magnitude of~\eqref{eq:energies_grav} with~\eqref{eq:energies_gp} gives the scaling $M_s \sim (\mpl/m)^2/r_s$. Using the numerically obtained curves (upper left corner of the left panel of fig.~\ref{fig:HandRvsN}), we have
\begin{align}\label{eq:Mh_grav}
    M_s \simeq & \left(\frac{10\,\mathrm{eV}}{m}\right)^2\left(\frac{10^{7}\,R_{\odot}}{r_{s,95}}\right)\,5.8\times 10^{-4}\,\mathrm{kg}.
\end{align}
For our benchmark value of $m = 10$ eV, such solitons are dilute (macroscopically large and low mass) and are very long lived (see~\ref{sec:pert_decay} ahead). Since dilute, they carry low amounts of intrinsic spin and/or iso-spin
\begin{align}
    \{|{\bm S}_{\mathrm{tot}}| \,,\,I\} \simeq 3.3 \times 10^{31}\left(\frac{10\,\mathrm{eV}}{m}\right)^3\left(\frac{10^{7}\,R_{\odot}}{r_{s,95}}\right)\,.
\end{align}
These objects lie towards the upper left corner of the left panel of Fig.~\ref{fig:HandRvsN}. 
Fig.~\ref{fig:field_profiles1} shows the scalar field $\psi$ and Newtonian potential $\Phi$ for a soliton in this regime, obtained numerically by shooting method. For solitons in this regime, using~\eqref{eq:delHdelN_mu} easily gives $M_{s} \propto \sqrt{\mu} \propto r_{s,95}^{-1}$. As $M_s$ (or equivalently $\mu$) increases, self interactions start to become important. Depending upon whether they are attractive or repulsive, we have drastically different behavior.\\

\noindent\textbf{Regime II: Attractive self-interactions balancing gradient pressure.} This is the case where $\kappa_{ym} = 0$, giving 
\begin{align}\label{eq:kappa_regime2}
    \kappa_{\mathrm{eff}} = -\frac{2m^2}{M_{\varphi}^2}\,\kappa_{\rm{h}},
\end{align}
and $\kappa_{h} \sim \mathcal{O}(1)$. In the upper half of table~\ref{tab:SU(2)_solitons}, we show solitons in the SU($2$) HYM which carry attractive self-interactions, and correspond to blue and red curves in fig.~\ref{fig:HandRvsN}; with the replacement $g \rightarrow g\sqrt{2}m/M_{\varphi}$. Starting from the previous regime, as $M_s$ increases, attractive self interactions start to become important and ultimately become comparable to gravity and gradient pressure. Balancing~\eqref{eq:energies_gp}, the magnitude of~\eqref{eq:energies_grav}, and the magnitude of~\eqref{eq:energies_self} against one another, and reading the numerical coefficient from fig.~\ref{fig:HandRvsN} (at the junction where dashed and solid blue/red curves meet), gives the following estimate for the mass
\begin{align}\label{eq:Ms_attr_trans}
    M_s|_{\mathrm{tr}} \simeq &\; \kappa_h^{-1/2}\left(\frac{g_{max}}{g}\right)\left(\frac{10\,\mathrm{eV}}{m}\right)^{7/4}\left(\frac{M_{\varphi}}{\mathrm{MeV}}\right)\times\nonumber\\
    &\,4.7 \times 10^{2}\,\mathrm{kg}\,,
\end{align}
and the associated radius
\begin{align}\label{eq:rs_attr_trans}
    r_{s,95}|_{\mathrm{tr}} \simeq &\, \kappa_h^{1/2}\left(\frac{g}{g_{max}}\right)\left(\frac{10\,\mathrm{eV}}{m}\right)^{1/4}\left(\frac{\mathrm{MeV}}{M_{\varphi}}\right)\,6.6\,R_{\odot}
\end{align}
at this transition point.\footnote{Upon including post Newtonian corrections, it is expected that the transition happens at slightly smaller values of $M_s$ (depending on the value of the gauge coupling $g$). See~\cite{Salehian:2021khb} for the scalar case with attractive self-interactions.} Upon comparison with~\eqref{eq:Mh_grav}, we see that there is a big window of macroscopically large solitons carrying not so small amounts of spin and/or iso-spin
\begin{align}
    \{|{\bm S}_{\mathrm{tot}}| \,,\,I\} \lesssim &\,\kappa_h^{-1/2}\left(\frac{g_{max}}{g}\right)\left(\frac{10\,\mathrm{eV}}{m}\right)^{11/4}\left(\frac{M_{\varphi}}{\mathrm{MeV}}\right)\times\nonumber\\
    &\,2.6 \times 10^{37}\,,
\end{align}
which only increase as $m$ decreases, and/or $M_{\varphi}$ increases. We shall see in ~\ref{sec:pert_decay} that solitons near this transition point are highly stable against perturbative decay due to number changing processes, and can therefore be very long lived.\\

\begin{figure}[!htb]
\begin{center}
\includegraphics[width=0.48\textwidth]{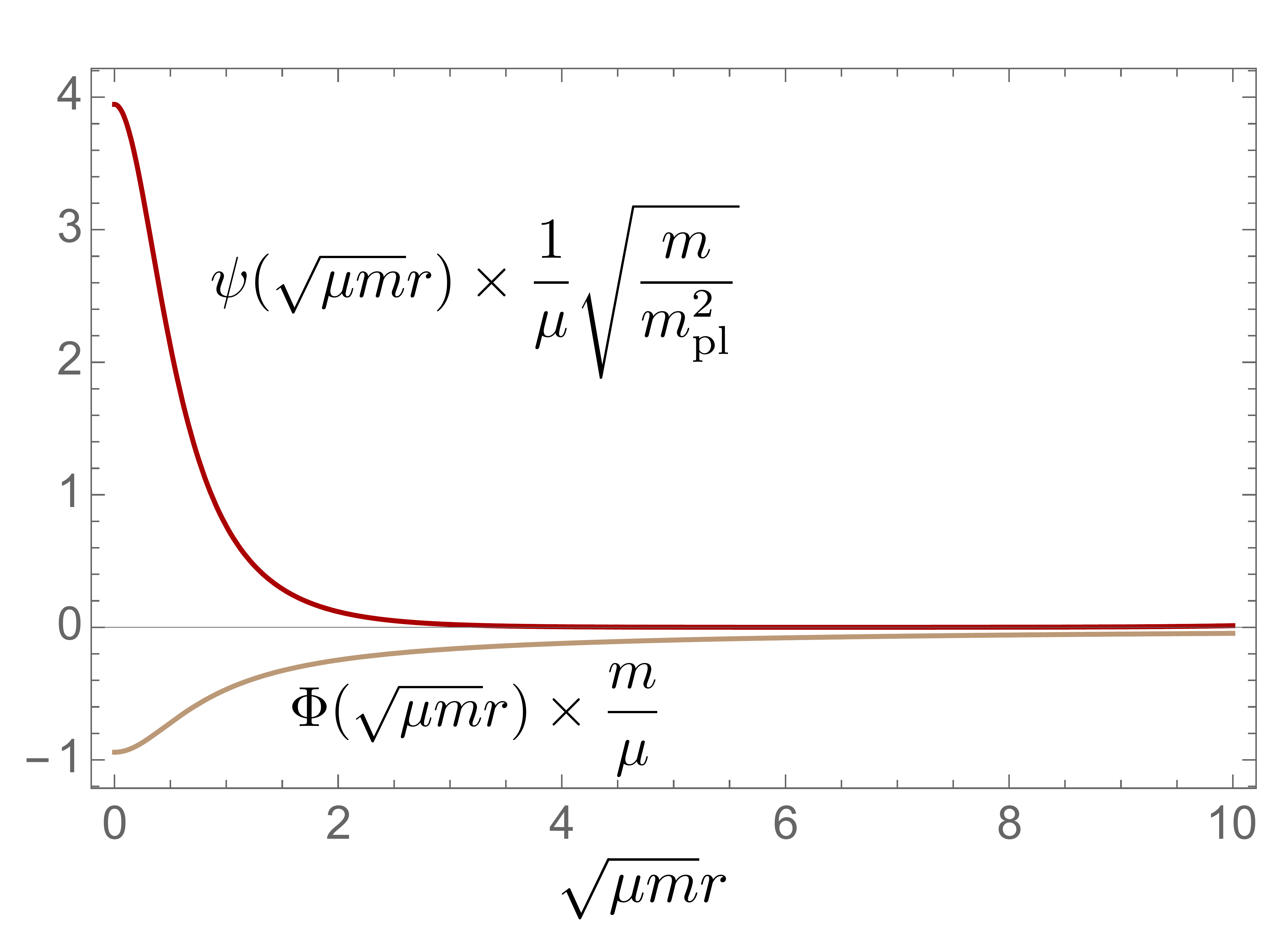}
\end{center}
\caption{Re-scaled scalar field profiles $\psi(r)$ and $\Phi(r)$ for a soliton 
held together primarily by a balance between attractive self interactions and gradient pressure. The numerical value of $\mu g^2\kappa_{\mathrm{h}}\mpl^2/(2mM_{\varphi}^2) = 0.624$ for the shown profile. This soliton lies on the dashed red/blue curves in fig.~\ref{fig:HandRvsN}, and is unstable.} 
\label{fig:field_profiles2}
\end{figure}

After this transition, we enter the regime 
where gravity starts to become sub-dominant, and attractive self-interactions balance the gradient pressure. Balancing~\eqref{eq:energies_gp} with~\eqref{eq:energies_self} gives $M_s \sim (M_{\varphi}/g)^2r_s/\kappa_h$. Once again using the numerically obtained curves (dashed curves on the lower left corner of the left panel of fig.~\ref{fig:HandRvsN}),
gives the following scaling
\begin{align}
    M_{s} \simeq &\; \kappa_h^{-1}\left(\frac{g_{max}}{g}\right)^2\left(\frac{10\,\mathrm{eV}}{m}\right)^{3/2}\left(\frac{M_{\varphi}}{\mathrm{MeV}}\right)^{2}\times\nonumber\\
    &\,\left(\frac{r_{s,95}}{\mathrm{km}}\right)\,2.8\times 10^{-4}\,\mathrm{kg}.\nonumber
\end{align}
Such solutions are however unstable against classical perturbations and lie on the dashed blue and red curves in fig.~\ref{fig:HandRvsN}. This instability can be easily observed by noting that the sign of $H_{\rm self}$ is negative, and corresponds to a local maximum in $H$ vs $r_{s,95}$ curve for a fixed mass~\cite{Schiappacasse:2017ham}. As such, these solitons are of little importance to us. For completeness however, we show the field profiles $\psi$ and $\Phi$ for a soliton in this regime in fig.~\ref{fig:field_profiles2} (obtained numerically by shooting method). Also, using the relationship~\eqref{eq:delHdelN_mu}, we get $M_{s} \propto \mu^{-1/2} \propto r_{s,95}$. This verifies that $\mu$ indeed increases as we traverse down the curves after the turn over/transition point.\\

Finally, we have the case when the self interactions are repulsive ($\kappa_{\mathrm{eff}} > 0$). These are the most interesting objects for our pivotal value of $m = 10$ eV, for they can be astrophysically large. We discuss them next.\\

\noindent\textbf{Regime III: Gravity balancing repulsive self interactions.} This is the case when $\kappa_{\mathrm{ym}} \neq 0$, and since $m^2/M_{\varphi}^2 \ll 1$, we have 
\begin{align}\label{eq:kappa_regime3}
    \kappa_{\mathrm{eff}} \approx \kappa_{\mathrm{ym}} \sim \mathcal{O}(1).
\end{align}
Once again starting from regime I, as $M_s$ increases, self interactions start to become important and we get to a transition point where gradient pressure, gravity, and repulsive self-interactions, all become comparable. To get an estimate of this transition point, we can once again balance all the three energy terms (Eq.~\eqref{eq:energies_gp}, the magnitude of~\eqref{eq:energies_grav}, and~\eqref{eq:energies_self} along with the relation~\eqref{eq:kappa_regime3}) against each other to obtain the following transition point mass
\begin{align}\label{eq:Ms_critical_Chandregime}
    M'_s|_{\mathrm{tr}} \sim \kappa_{\mathrm{ym}}^{-1/2}\left(\frac{g_{max}}{g}\right)\left(\frac{10\,\mathrm{eV}}{m}\right)^{3/4}\,\mathcal{O}(10^{-2}-10^{-3})\,.
\end{align}
For masses much larger than this, we end up in the regime where gradient pressure becomes sub-dominant, while gravity balances the repulsive self-interactions. In this regime, the radii of solitons start to become independent of their mass, as can be seen by balancing the magnitude of~\eqref{eq:energies_grav} against~\eqref{eq:energies_self}. We get
\begin{align}\label{eq:rs_bound}
    r_{s,95} \simeq \frac{\kappa_{\mathrm{ym}}^{1/2}\,g\,\mpl}{m^2} \simeq &\, \kappa^{1/2}_{\mathrm{ym}}\left(\frac{g}{g_{\mathrm{max}}}\right)\left(\frac{10\,\mathrm{eV}}{m}\right)^{5/4}\times\nonumber\\
    &\,2\times 10^5\,R_{\odot}\,,
\end{align}
dictating astrophysically large solitons.\\

Numerically, we observe a very weak decreasing trend in $r_s$ as $M_s$ increases to larger and larger values (see the left panel of~\ref{fig:HandRvsN2}), consistent with what was reported in~\cite{Schiappacasse:2017ham}. For our purposes in this paper, we neglect this mild dependence. 
Upon inclusion of GR effects, the curve is ultimately expected to turn over as found in~\cite{Croon:2018ybs,Salehian:2021khb} for scalar solitons with repulsive self-interactions. This however, only at very large values of $M_s$ when $GM_s \sim r_s$. Requiring $M_s \lesssim r_s/G$ in order for GR effects to be negligible and our non-relativistic treatment to hold, \textit{together} with~\eqref{eq:rs_bound}, fetches the following upper bound on the mass
\begin{align}\label{eq:Ms_upper}
    M_s \lesssim \kappa_{\mathrm{ym}}^{1/2}\left(\frac{g}{g_{max}}\right)\left(\frac{10\,\mathrm{eV}}{m}\right)^{5/4}\,10^{11}\,M_{\odot}.
\end{align}
Such massive solitons may turn out to be unstable owing to quantum mechanical effects. In section~\ref{sec:pert_decay} we will see that solitons in this regime, with masses as big as $M_s \sim 10 M_{\odot}$ (for our benchmark values of $m$ and $g$), can still have lifetimes larger than the age of the Universe. For the polarization, it is the $\kappa_{\ym} = 1$ family of solitons that carry both spin and iso-spin, and can be astrophysical
\begin{align}
    |{\bm S}_{\mathrm{tot}}| = I \simeq 10^{66}\,\left(\frac{M_s}{10\,M_{\odot}}\right)\left(\frac{10\,\mathrm{eV}}{m}\right)\,.
\end{align}
Note that the scaling of the upper bound on the mass (c.f.~\eqref{eq:Ms_upper}) is the same as Chandrasekhar mass limit $M_s \propto \mpl^3\,m^{-2}$ (using $g_{\mathrm{max}} \propto m^{3/4})$. This is understood on account of the fact that gravity is being balanced by a repulsive force (albeit due to Yang-Mills self interaction, as opposed to Pauli exclusion in fermions). Owing to this behavior, we call solitons in this regime as `Chandrasekhar solitons'. Lastly, for this regime to exist in the first place, we want the upper bound on $M_s$ (given by~\eqref{eq:Ms_upper}) to be larger than $M'_s|_{\mathrm{tr}}$~\eqref{eq:Ms_critical_Chandregime}. This gives the following estimate for the lower bound on the gauge coupling
\begin{align}
    g \gtrsim \kappa_{\mathrm{ym}}^{-1/2}\left(\frac{m}{10\,\mathrm{eV}}\right)\mathcal{O}(10^{-26}-10^{-27})\,.
\end{align}
Together with the Bullet cluster bound~\eqref{eq:g_bound}, this provides for a large window for the gauge coupling, giving rise to a large range for Chandrasekhar solitons.\\

In this regime, it becomes difficult to numerically obtain soliton solutions for very large values of $M_s$. However, since we now know the behaviour of $M_s$ and $r_{s,95}$, we can extrapolate from some point (on the right panel of fig.~\ref{fig:HandRvsN2}) to get to more massive solitons. Using $H \propto -M_s^2$ in this regime and beginning from $M_s \simeq 2\times 10^3\,\mpl/g$ for $\kappa_{\ym} = 1$,\footnote{This corresponds to $\mu/m = 7.5\,m^2/(g^2\mpl^2)$.} we get the following estimate for the energy of the Chandrasekhar solitons
\begin{align}\label{eq:HvsM_extrapolate}
    \frac{H}{M_{\odot}} \sim & - 10^{-11}\,\kappa_{\ym}^{-1/2}\left(\frac{g_{max}}{g}\right)\left(\frac{m}{10\,\mathrm{eV}}\right)^{5/4}\times\nonumber\\
    &\,\left(\frac{M_s}{10\,M_{\odot}}\right)^2
\end{align}
for larger soliton masses. Once again using~\eqref{eq:delHdelN_mu}, we see $M_{s} \propto \mu$. With other parameters at their benchmark values, $\mu/m \sim 10^{-12}(M_s/10\,M_{\odot})$, indictating that we are well within the non-relativistic regime.\\

Note that since we assumed the scaling~\eqref{eq:rs_bound}, we have neglected the weak dependence of $r_s$ on $M_s$ as stated earlier. A full GR analysis is required to capture the corrections, which we leave for future work.\\

\begin{figure}[!htb]
\begin{center}
\includegraphics[width=0.48\textwidth]{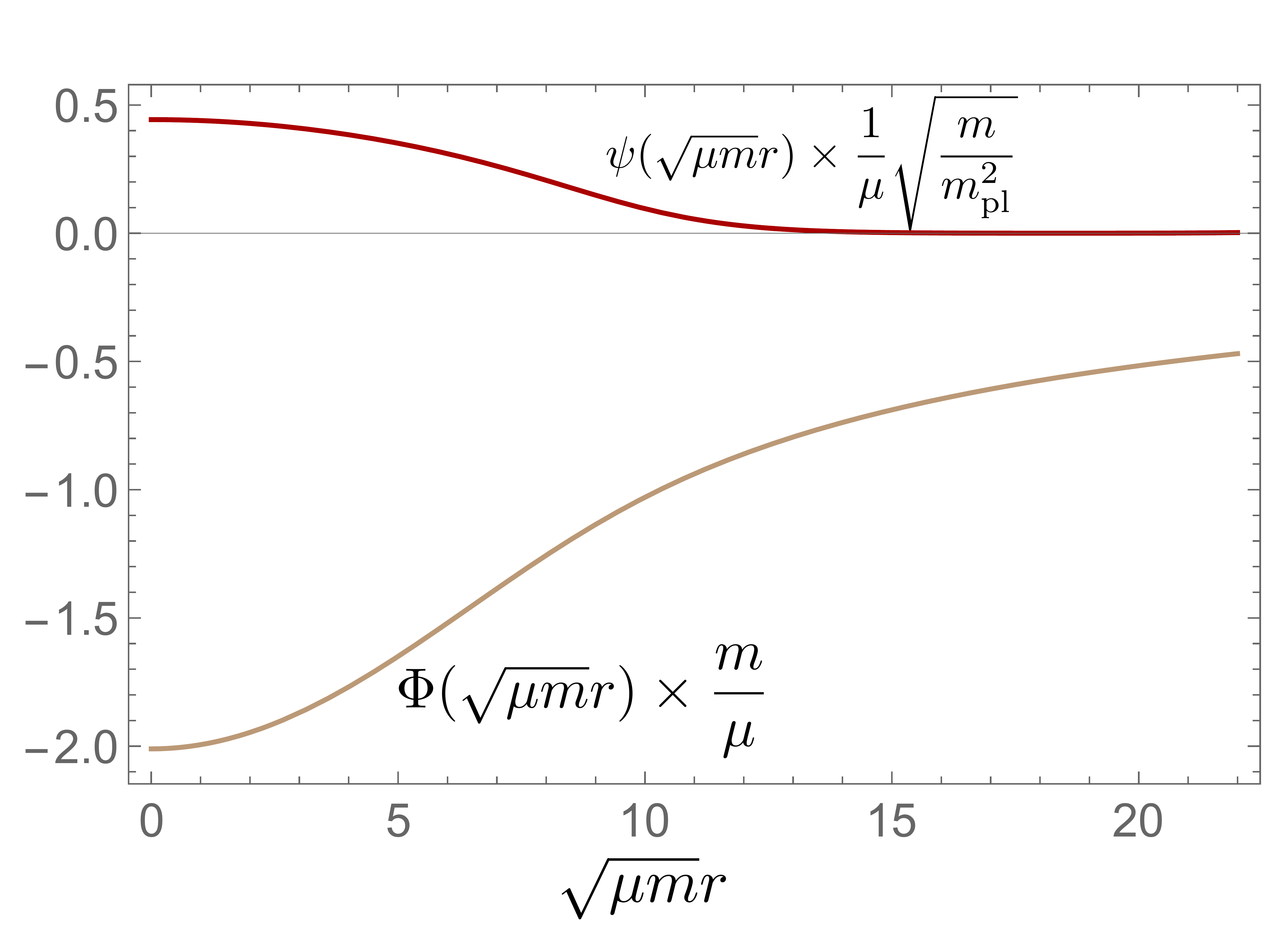}
\caption{Re-scaled scalar field profiles $\psi(r)$ and $\Phi(r)$ for a Chandrasekhar soliton 
(held together by a balance between gravity and repulsive self-interactions, with gradient pressure being only sub-dominant). The numerical value for $\mu g^2\kappa_{\mathrm{ym}}\mpl^2/(4m^3) = 5$. Depending upon the value of $\kappa_{\rm ym}$ (for a given set of $\mu$, $g$, and $m$), the corresponding soliton falls on one of the four curves of fig.~\ref{fig:HandRvsN2}. For $\kappa_{\rm ym} = 1$ (orange curve in fig.~\ref{fig:HandRvsN2}), the soliton carries both intrinsic spin and iso-spin, and is pictorially represented in fig.~\ref{fig:example_soliton}}
\label{fig:field_profiles3}
\end{center}
\end{figure}

\begin{figure*}[t]
\begin{center}
\includegraphics[width=0.95\textwidth]{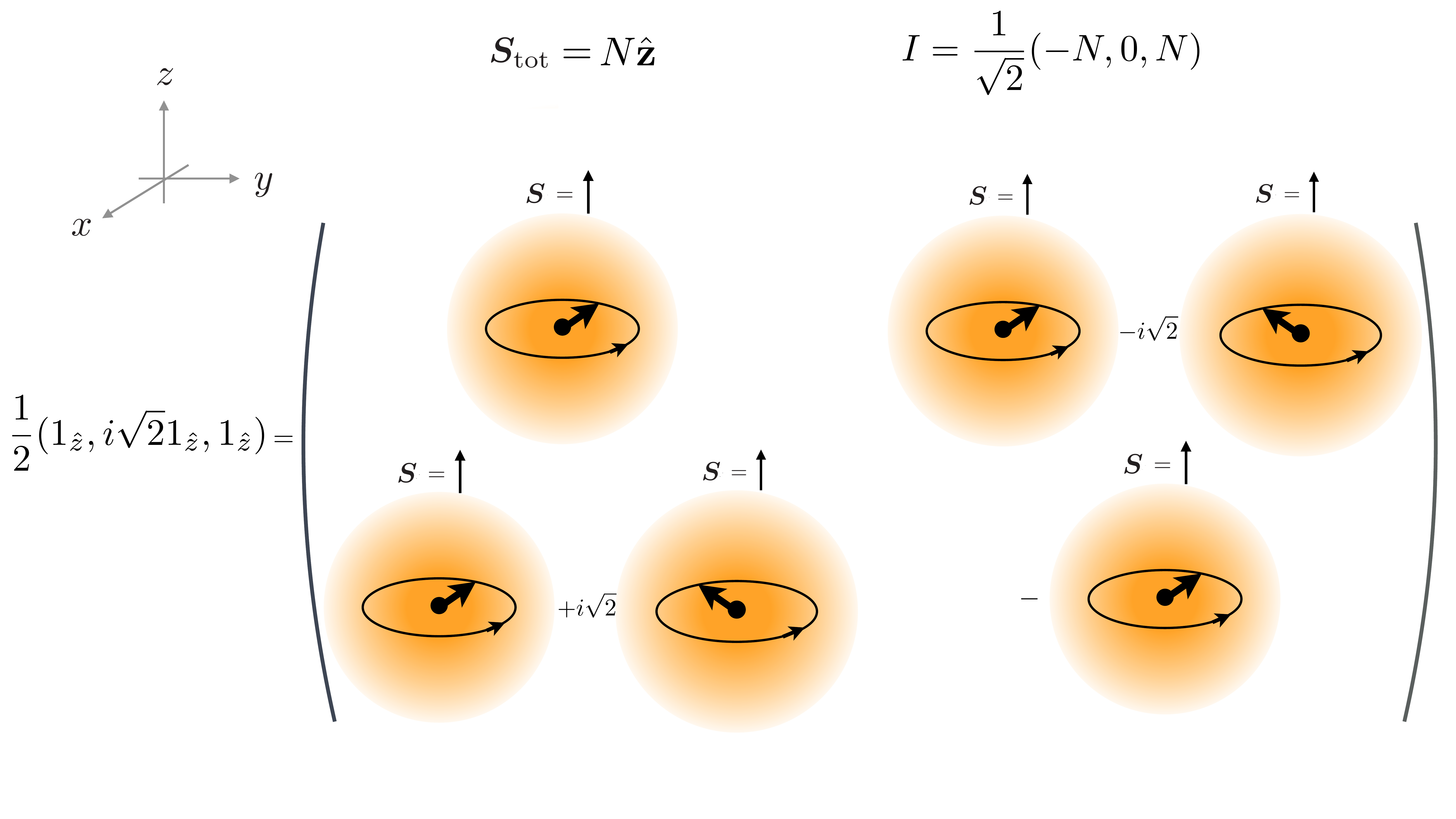}
\caption{A pictorial representation of the $\frac{1}{2}(1_{\hat{z}},i\sqrt{2}1_{\hat{z}},1_{\hat{z}})$ soliton given by~\eqref{eq:sample_soliton}. The orange balls represent the radially symmetric profile $\psi(r)$, with the total number of particles $N = \int \mathrm{d}^3x \psi^2(r)$. Within the orange balls, the big arrows represent the vector fields ${\bm W}_a$, while the circles with small arrows represent the time evolution of these vector fields. Note that there is a $\pi/2$ phase difference between ${\bm W}_2$ and ${\bm W}_3$, which results in an iso-spin density $\frac{1}{\sqrt{2}}(-\psi^2,0,\psi^2)$ and hence the total iso-spin $I = \frac{1}{\sqrt{2}}(-N,0,N)$. The soliton also has a spin density $\bS = \psi^2\hat{\bm z}$ and correspondingly the total spin $|\bS_{\mathrm{tot}}|= N$. Depending upon $N$, these quantities can be astrophysically large.} 
\label{fig:example_soliton}
\end{center}
\end{figure*}

Second set in Table~\ref{tab:SU(2)_solitons} shows solitons with repulsive self-interactions (in the SU($2$) HYM theory). As a concrete example, consider the $\frac{1}{2}(1_{\hat{z}},i\sqrt{2}1_{\hat{z}},1_{\hat{z}})$ soliton (for which $\kappa_{\mathrm{ym}} = 1$). Using~\eqref{eq:W_decompose}, the full Yang-Mills field ${\bm W} = {\bm W}^a\tau_a$ (in the non-relativistic approximation) is
\begin{widetext}
\begin{align}\label{eq:sample_soliton}
    {\bm W}(r,t) \approx \frac{1}{2\sqrt{2m}}\psi(r)\begin{pmatrix}
    \begin{pmatrix} \cos \omega t\\ \sin \omega t \\ 0\end{pmatrix} & \begin{pmatrix} \cos \omega t\\ \sin \omega t \\ 0\end{pmatrix} - i\sqrt{2}\begin{pmatrix} \sin \omega t\\ -\cos \omega t \\ 0\end{pmatrix}\\
    \begin{pmatrix} \cos \omega t\\ \sin \omega t \\ 0\end{pmatrix} + i\sqrt{2}\begin{pmatrix} \sin \omega t\\ -\cos \omega t \\ 0\end{pmatrix} & -\begin{pmatrix} \cos \omega t\\ \sin \omega t \\ 0\end{pmatrix}\\
    \end{pmatrix}
\end{align}
\end{widetext}
where $\omega = m-\mu$, and we have assumed the (commonly used) representation in which the third Pauli matrix $\tau_3$ is diagonal.\footnote{In this representation, the three Pauli matrices are $\tau_1 = \begin{pmatrix}0 & 1\\ 1& 0\end{pmatrix}$, $\tau_2 = \begin{pmatrix}0 & -i\\ i& 0\end{pmatrix}$, and $\tau_3 = \begin{pmatrix}1 & 0\\ 0& -1\end{pmatrix}$.} The total particle number is $N = \int\mathrm{d}^3x\,\psi^2 = M_s/m$, with
\begin{align}\label{eq:spin.n.iso-spin.Chand.sol}
    {\bm S}_{\mathrm{tot}} &= \frac{M_s}{m}\,\hat{\bm z}\,,\nonumber\\
    I &= \frac{1}{\sqrt{2}}(-1,0,1)\frac{M_s}{m}\,.
\end{align}
The field profile $\psi(r)$ and the Newtonian potential $\Phi(r)$ of such a soliton configuration, i.e. $\kappa_{\rm ym} = 1$, for $\mu/m = 20m^2/(g^2\mpl^2)$, is the same as that shown in Fig.~\ref{fig:field_profiles3}. Fig.~\ref{fig:example_soliton} shows a pictorial representation of a $\frac{1}{2}(1_{\hat{z}},i\sqrt{2}1_{\hat{z}},1_{\hat{z}})$ soliton.

\subsection{Perturbative decay and lifetimes}\label{sec:pert_decay}

Quantum mechanical effects have been speculated to lead to perturbative decay of solitons~\cite{Hertzberg:2010yz,Hertzberg:2020xdn}. In particular, short wavelength perturbations (of the order of the mass $m$ of the bosonic particle) are expected to emerge within a soliton solution, leading to number changing processes on account of self-interactions, and hence soliton decay. Having discussed the various possible solitons in the SU($2$) HYM theory, in this section we provide estimates of their lifetimes due to such particle annihilation processes within them. To leading order in the non-relativistic limit, the $4$-point vertex in the Yang-Mills potential~\eqref{eq:Vhym} suffices, and gives rise to a $4 \rightarrow 2$ process. The tree-level decay rate due to such a process can be estimated to be~\cite{Hertzberg:2010yz,Hertzberg:2020xdn}
\begin{align}\label{eq:decay_est}
    \Gamma_{4\rightarrow 2} \sim \frac{|\mathcal{M}_{s}|^2\,k_{\mathrm{rad}}\int\mathrm{d}^3x\,n^4_s}{m^5\int\mathrm{d}^3x\,n_s}\,,
\end{align}
where $|\mathcal{M}_s| \sim \kappa_{\mathrm{eff}}^2\,g^4/m^2$ is the matrix element associated with the $4 \rightarrow 2$ process, $k_{\mathrm{rad}} \sim m$ is the momentum carried by the outgoing/produced particles, and $n_s = \psi^2(r)$ is the number density of the soliton. The quantity $\int\mathrm{d}^3x\,n^4_s/\int\mathrm{d}^3x\,n_s \sim M_s^3/(m^3\,r_s^9)$.\footnote{Numerically, we find $\int\mathrm{d}^3x\,n^4_s/\int\mathrm{d}^3x\,n_s = f\,M_s^3/(m^3\,r_s^9)$ with $f \rightarrow 1$ for when the self interactions become less important and the soliton is maintained by a balance between gravity and gradient pressure (regime I). On the other hand for Chandrasekhar solitons in regime III, extrapolation to larger values of $M_s$ suggests $f \rightarrow 0.04$.} Then, the lifetime of an isolated soliton can be defined as the inverse of this decay rate $\tau \equiv 1/\Gamma_{4\rightarrow 2}$:
\begin{align}
    \tau = \alpha\,\frac{m^{11}r_s^9}{g^8\,M_s^3\,\kappa^4_{\mathrm{eff}}}\,,
\end{align}
where $\alpha$ is a cumulative factor resulting from keeping track of all the order unity factors in the above scalings. With this, we can now estimate the lifetimes of solitons that lie on the stable branches of the two different families (solid curves in Fig.~\ref{fig:HandRvsN} having negative energy $H$).\\

\noindent\textbf{Attractive self interactions:} For solitons with attractive interactions in the core ($\kappa_{ym} = 0$ and $|\kappa_{\mathrm{eff}}| = 2\,\kappa_h\,m^2/M_{\varphi}^2$) that lie in regime I, using~\eqref{eq:Mh_grav} gives the following estimate for their lifetime
\begin{align}\label{eq:lifetime1}
    H_0\tau \simeq &\; 4 \times 10^{191}\,\alpha\,h\,\kappa_h^{-4}\left(\frac{g_{max}}{g}\right)^{8}\left(\frac{10^{-6}\,\mathrm{kg}}{M_s}\right)^{12}\times\nonumber\\
    &\,\left(\frac{10\,\mathrm{eV}}{m}\right)^{21}\left(\frac{M_{\varphi}}{\mathrm{MeV}}\right)^{8}\,,
\end{align}
dictating tremendous longevity. We have used $H_0 = 100\,h\,$ km/Mpc/s. Furthermore, the transition point up until which such solitons are stable against classical perturbations, correspond to having $M_s$ and $r_s$ given by~\eqref{eq:Ms_attr_trans} and~\eqref{eq:rs_attr_trans} respectively. For solitons near this transition point, we get
\begin{align}\label{eq:lifetime1.1}
    H_0\tau|_{\mathrm{tr}} \simeq &\; 10^{85}\,\alpha\,h\,\kappa_h^{-2}\left(\frac{g}{g_{max}}\right)^{4}\left(\frac{\mathrm{MeV}}{M_{\varphi}}\right)^{4}\,,
\end{align}
which dictates that the full (classically stable) branch of solitons hosting attractive self-interactions within them, correspond to very long lived solitons (against these perturbative decays).\\

\noindent\textbf{Repulsive self-interactions:} For solitons with repulsive self-interactions within them, there are two regimes of interest. First (like the previous case) is regime I where gravity is dominant and balances the gradient pressure. Using~\eqref{eq:Mh_grav} with $\kappa_{\mathrm{eff}} = \kappa_{\mathrm{ym}}$, we get
\begin{align}\label{eq:lifetime2.1}
    H_0\tau \simeq &\; 5.7\times 10^{152}\,\alpha\,h\,\kappa^{-4}_\mathrm{ym}\left(\frac{g_{max}}{g}\right)^{8}\left(\frac{10^{-6}\,\mathrm{kg}}{M_s}\right)^{12}\times\nonumber\\
    &\,\left(\frac{10\,\mathrm{eV}}{m}\right)^{13},
\end{align}
once again dictating tremendously long lived solitons.\\

Most importantly, for Chandrasekhar solitons in regime III (with radii approximately independent of their mass~\eqref{eq:rs_bound}), we get
\begin{align}\label{eq:lifetime2.2}
    H_0\tau \simeq &\, 13\,\alpha\,h\,\kappa^{1/2}_{\mathrm{ym}}\left(\frac{g}{g_{\mathrm{max}}}\right)\left(\frac{10\,M_{\odot}}{M_s}\right)^3\times\nonumber\\
    &\,\left(\frac{10\,\mathrm{eV}}{m}\right)^{25/4}\,,
\end{align}
dictating cosmological lifetimes for $r_{s} \sim 10^{5} R_{\odot}$ solitons that are as massive as $10 M_{\odot}$. These are the most interesting objects for our benchmark values of $m = 10$ eV and $g \simeq g_{\mathrm{max}}$. Note the severe dependence (high exponents) of these lifetimes on different parameters $m$, $M_{\varphi}$, $M_{s}$, and $g$. Especially the mass $m$ for Chandrasekhar solitons. Decreasing $m$ by a factor of $10$ makes the lifetime increase by a factor of about $10^6$. So it becomes relatively easy to get to higher lifetimes (albeit at the expense of smaller gauge couplings due to Bullet cluster bounds~\eqref{eq:g_bound}). To get precise/concrete results for the lifetimes (including the mild dependence of $r_s$ on $M_s$ in the latter case above), a fully relativistic treatment, including 3-D simulations, is warranted. We shall pursue this exercise in a separate work.

\subsection{Parametric resonance}\label{sec:resonance_decay}

Apart from perturbative decay, solitons may also be susceptible to instabilities due to the phenomenon of parametric resonance. Since the field oscillates coherently, certain frequency perturbations can experience exponential growth due to them having a time dependent (periodic) effective mass. The relevant quantity is the second derivative of the potential, evaluated with the background soliton solution. Near the core of the soliton, one may approximate the background as being homogeneous, in order to estimate the behavior of different frequency perturbations. Then, the condition to avoid resonance can be formulated as $\mu_k\,r_{s} \lesssim 1$ where $\mu_k$ are the (frequency dependent) Floquet exponents. See~\cite{Hertzberg:2018zte,Hertzberg:2014jza,Hertzberg:2020xdn} for a detailed analysis of the scalar case.\\

The case with vector fields is different, because of the non-trivial structure of the interaction potential and the fact that soliton configurations can have spin and/or iso-spin. The second derivative of the potential was necessarily time dependent for the scalar case since the potential was time dependent (periodic in time). For the vector case however, configurations with spin and/or iso-spin have a \textit{time independent} potential term in the leading non-relativistic limit. That is $V = V_{\mathrm{ym}} + V_{\mathrm{h}}\propto \psi^4(r)$. These are the $(0_{\hat{z}},i\sqrt{2}\,0_{\hat{z}},0_{\hat{z}})/2$, and $(1_{\hat{z}},1_{\hat{z}},1_{\hat{z}})/\sqrt{3}$ solitons for the case of attractive self-interaction; while $(1_{\hat{z}},i\sqrt{2}\,1_{\hat{z}},1_{\hat{z}})/2$ for the case of repulsive self-interaction. 
The relevant quantity for the equation of motion of vector perturbations is the Hessian matrix, given by the sum of the second derivative of the Yang-Mills potential
\begin{align}
    \frac{\partial^2 V_{\ym}}{\partial W^a_{i}\,\partial W^b_{j}} =&\,g^2\Biggl(2\,W_{a}^{i}W_{b}^{j} + W_{c}^{k}W^{c}_{k}\,\delta_{ab}\,\delta^{ij} - W_{c}^{i}W^{j}_{c}\,\delta_{ab}\nonumber\\
    &\quad - W_{a}^{j}W_{b}^{i} - W^{a}_{k}W_{b}^{k}\,\delta^{ij}\Biggr)\,,
\end{align}
and the Higgs induced potential
\begin{align}
    \frac{\partial^2 V_{\mathrm{h}}}{\partial W^a_{i}\,\partial W^b_{j}} =&\, - \frac{g^2m^2}{M_{\varphi}^2}\Biggl(2\,W_{a}^{i}W_{b}^{j} + W_{c}^{k}W^{c}_{k}\,\delta^{a}_{b}\,\delta^{ij}\Biggr)\,.
\end{align}
Here we have discarded the zeroth components of the vector fields in the leading non-relativistic limit. We construct a $9 \times 1$ vector $({\bm W}^1, {\bm W}^2, {\bm W}^3)$, where each ${\bm W}^a = (W^a_x, W^a_y, W^a_z)$. Correspondingly we write down a $9 \times 9$ Hessian matrix from the sum of the above two terms. The goal then, is to see if it has any time-dependent eigenvalues. If yes, then the corresponding eigenmode may grow due to parametric resonance, otherwise not. For the above mentioned polarized solitons, we find that all the $9$ eigenvalues are \textit{time-independent} (and of-course proportional to the square of the soliton profile $\psi^2(r)$). This implies that polarized solitons (having non-zero spin and/or iso-spin) are safe against this instability due to parametric resonance.\\

As already mentioned, but nevertheless important, the above result holds in the leading non-relativistic limit. This may or may not change upon inclusion of relativistic corrections. We leave a detailed analysis with the inclusion of relativistic corrections for future work.

\section{U($1$) Abelian Higgs as a special case}\label{sec:Abelian.special.case}

Having discussed the HYM SU($2$) case, in this section we quickly discuss results for the special case of U($1$) Abelian Higgs model. Since the Yang-Mills self interactions are absent ($V_{\ym} = 0$), only the Higgs induced attractive ones are active. The two soliton solutions (with zero orbital angular momentum) possible would be the first and third rows in table~\ref{tab:SU(2)_solitons}, describing linearly polarized and circularly polarized solitons respectively. This can be seen by performing an (iso-)rotation in the internal space to give $(0_{\hat{z}},0_{\hat{z}},0_{\hat{z}})/\sqrt{3} \rightarrow (0_{\hat{z}},\times,\times)$, and $(1_{\hat{z}},1_{\hat{z}},1_{\hat{z}})/\sqrt{3} \rightarrow (1_{\hat{z}},\times,\times)$, where `$\times$' corresponds to a null entry.\\

To read out the Bullet cluster constraint in this case, first we replace $g \rightarrow \sqrt{2}\,g\,m/M_{\varphi}$ in~\eqref{eq:cross_section}.\footnote{This is because the ratio of the coefficients of $V_{\mathrm{h}, \mathrm{nr}}$ and $V_{\ym, \mathrm{nr}}$ is $2m^2/M_{\varphi}^2$ (equivalently, $\kappa_{\mathrm{h}}/\kappa_{\ym} = 2m^2/M_{\varphi}^2$) and they are proportional to $g^2$.} Then, using $M_{\varphi} = 2\sqrt{2\lambda}m/g$, we get $\sigma = g^8|\mathcal{A}|^2/(1024\pi \lambda^2m^2)$. Following a similar logic of equipartition, $|\mathcal{A}|$ can be estimated as $|\mathcal{A}| = 13/3$ to give the following bound
\begin{align}\label{eq:g_bound_U1}
    g \lesssim 5.4\times 10^{-3}\,\eta^{1/8}\,\left(\frac{m}{10\,\mathrm{eV}}\right)^{3/8}\lambda^{1/4} \equiv g'_{max}\,.
\end{align}
Since the allowed values of couplings are larger, the transition point after which solitons start to become classically unstable; dotted red/blue curves in fig.~\ref{fig:HandRvsN}; moves towards smaller (larger) values
of $M_s$ ($r_{s,95}$). We get the following estimates
\begin{align}\label{eq:Ms.rs_attr_trans_U1}
    M_s|_{\mathrm{tr}} \simeq &\; \kappa_h^{-1/2}\left(\frac{g'_{max}}{g}\right)^2\left(\frac{10\,\mathrm{eV}}{m}\right)^{3/4}\,1.5 \times 10^{-2}\,\mathrm{kg}\nonumber\\
    r_{s,95}|_{\mathrm{tr}} \simeq &\, \kappa_h^{1/2}\left(\frac{g}{g'_{max}}\right)^2\left(\frac{10\,\mathrm{eV}}{m}\right)^{5/4}\,2\times 10^{5}\,R_{\odot}\,.
\end{align}
Since there is no Chandrasekhar branch in this case, the mass of the vector field $m$ must be very small in order for the stable solitons (supported by a balance between gravity and gradient pressure) to be astrophysical. The corresponding values of the gauge coupling are much smaller as compared to the Yang-Mills case. From~\eqref{eq:lifetime1} and~\eqref{eq:lifetime1.1}, lifetimes can be straightforwardly obtained by normalizing $g$ with $g'_{\mathrm{max}}$, while also using $M_{\varphi} = 2\sqrt{2\lambda}m/g$.

\section{Summary and Future directions}
\label{Sec:sum_disc}

With the specific example of SU($2$), in this work we explored various non-topological solitons present in the Higgs phase of a non-Abelian dark sector. We first derived the effective Higgsed Yang-Mills (HYM) theory of massive spin-1 dof, by integrating out the heavy Higgs. Furthermore, with much lower ambient energy scales in the dark sector (much less than the mass $m$ of the vector fields), we have an effective non-relativistic HYM theory at our disposal. We derived this effective theory, and discussed all the different symmetries and the associated conserved quantities.\\ 

While the Higgs induced quartic interaction between the massive vector fields is attractive, the inherent Yang-Mills quartic-self interaction on the other hand (which is the relevant interaction in the non-relativistic limit of the theory), is repulsive. Due to the presence of both such interactions alongside gravity, the theory admits all possible types of soliton solutions, that we call `Yang-Mills stars'. Depending upon the kind of interaction (attractive or repulsive) that a given Yang-Mills star hosts, it can have varying mass and size, and be astrophysically large and long lived. Most importantly, due to the spin-$1$ nature of the massive vector fields, these Yang-Mills stars can carry huge amounts of intrinsic spin (similar to the case of a single massive vector field~\cite{Jain:2021pnk,Zhang:2021xxa}), and also iso-spin (due to the non-Abelian structure of the theory).\\ 

By estimating their perturbative decay rates (due to the leading $4 \rightarrow 2$ process within them), while also checking for parametric resonance growth of their perturbations, we found that Yang-Mills stars can be very long lived. Furthermore, stars with intrinsic spin/iso-spin may further be safe against exponential growth of perturbations due to parametric resonance. As such, this provides for interesting phenomenological avenues both in the early and late Universe. Table~\ref{tab:SU(2)_solitons} shows the diverse zoo of possible solitons in SU($2$) HYM theory.\\

For the presentation in this paper, we have normalized the mass of the vector fields by $m = 10$ eV, the gauge coupling by $g = g_{\mathrm{max}} \sim 10^{-5}$ (with $g_{\mathrm{max}}$ being the upper bound due to Bullet cluster constraints), and the dark Higgs mass by $M_{\varphi} = $ MeV. Following are some of the estimates for the sizes, masses, and spin/iso-spin of the various Yang-Mills stars, for when these parameters are at the aforementioned benchmark values. See Sec.~\ref{Sec:grav_bound} for general scalings.

Yang-Mills stars that are held together mostly by a balance between gravity and gradient pressure (regime I), are very dilute. For instance, $M_s \simeq 6$ kg for $r_{s,95} \simeq 10^{3}\,R_{\odot}$, with $M_s \propto 1/r_{s,95}$ in general. As $M_s$ increases, we get drastically different behavior depending upon whether the self-interactions are attractive or repulsive.

For the case of attractive self-interactions, the transition point at which they become comparable to gravity and gradient pressure, occurs around $M_{s}|_{\mathrm{tr}} \sim 470$ kg and $r_{s,95}|_{\mathrm{tr}} \sim 6\,R_{\odot}$, for which the solitons can carry $\sim \mathcal{O}(10^{37})$ amounts of spin or iso-spin. Such solitons are tremendously long lived, and even though very dilute, may nevertheless be interesting from a phenomenological point of view owing to such macroscopic spin/iso-spin. After this transition, gravity starts to become sub-dominant (regime II) and $M_s$ starts to decrease as $r_{s,95}$ decreases. Solitons in this regime are unstable and of no practical importance.

On the other hand for solitons hosting repulsive self-interactions, once the gradient pressure becomes sub-dominant and the soliton is held together by a balance between gravity and repulsive self-interaction, we get what we refer to as Chandrasekhar solitons (regime III). Here, the radius becomes (almost) independent of the mass of the soliton $r_s \sim g\,\mpl\,m^{-2}$ ($ \sim 10^{5} R_{\odot}$ for our benchmark values). With this, the maximum possible mass, obtained by requiring $M_s$ be less than the Schwarzchild mass for the above radius, has the same scaling as that for the Chandrasekhar mass limit $M_s \propto \mpl^3\,m^{-2}$; hence the name. Chandrasekhar solitons as massive as $\mathcal{O}(10) M_{\odot}$, can still have longer lifetimes than the age of the Universe, and can carry very large amounts ($\sim 10^{66}$) of intrinsic spin and iso-spin polarization.

\subsection{Future directions}\label{Sec:future_dir}

\noindent\textbf{Production mechanisms}: Production of massive spin-1 fields in the early Universe, that also serve as the observed dark matter, has been gaining much attention. Several authors have proposed different scenarios, although mostly in the context of massive spin-1 sector(s) \textit{without} any significant self-interactions. For instance, coupling the gauge kinetic term with a coherently oscillating (mis-aligned) scalar $\Delta\mathcal{L} \subset (a/f_a)\Tr{[\tilde{G}_{\mu\nu}G^{\mu\nu}]}$, can lead to efficient production of such fields~\cite{Agrawal:2018vin,Co:2018lka,Dror:2018pdh}. Gravitational particle production could be another mechanism to produce these fields in the very early Universe~\cite{Graham:2015rva,Kolb:2020fwh,Alexander:2020gmv,Sato:2022jya}. There may also exist resonant production scenarios during/after phase transition in the dark sector as outlined in~\cite{Nakayama:2021avl}. Introducing non-negligible self interactions usually leads to various caveats. A detailed analysis is required to better understand the fate of such fields in our Universe. This will be the subject of investigation in~\cite{Jain:2022}.\\

\noindent\textbf{Relativistic corrections}: In this work, we pursued solitonic solutions within the non-relativistic limit, and provided rough estimates for their decay rates and lifetimes (by considering $4 \rightarrow 2$ processes within their cores). However a detailed and a fully relativistic analysis is needed to analyze their stability and lifetimes accurately. Furthermore, gravitational waves could be sourced at higher orders (in Post Newtonian/relativistic corrections), which can lead to interesting spin dependent signatures. See~\cite{Helfer:2018vtq} for the case of scalar solitons (supported only by gravity).\\

\noindent\textbf{Early Universe phenomenology}: For mass values as `high' as $10$ eV, it is reasonable to think that our Yang-Mills starts (especially the astrophysical Chandrasekhar solitons) could be present in the early Universe, even earlier than the photon decoupling epoch. For instance if one considers the scenario of producing these vector fields through some misaligned scalar~\cite{Agrawal:2018vin}, the Hubble around the time of its production is also a few eV, which is much earlier than the CMB epoch. 
Upon multiplying~\eqref{eq:lifetime2.2} by $H_{\mathrm{cmb}}/H_0$, we see that Chandrasekhar solitons even as massive as $M_s \sim 10^3-10^4 M_{\odot}$, have longer lifetimes than $H_{\mathrm{cmb}}^{-1}$ (with other parameters set at their pivot values), making them relevant during that epoch. This can lead to interesting phenomenology and detection aspects, especially owing to their macroscopic spin and/or iso-spin. See~\cite{Bai:2020jfm} for instance, for the case of dark MACHOS present at the time of CMB.\\ 

Irrespective of their presence and abundance at such early times, such solitons might be present in our Contemporary Universe and can have other astrophysical implications as well~\cite{Blinov:2021axd}.\\

\noindent\textbf{Couplings with the Standard Model}: There could be several possibilities for couplings of the HYM dark sector with the SM. To name a few, one possibility could be to have higher dimensional operators (greater than $4$) in the low energy limit. Examples include axion-photon like couplings
\begin{align}
    \Delta\mathcal{L} \sim \frac{1}{M'^2}\Tr[W_{\mu}W^{\mu}]B_{\alpha\beta}\tilde{B}^{\alpha\beta}\,,
\end{align}
emerging due to some CP violating physics in the UV, or simply
\begin{align}
    \Delta\mathcal{L} \sim \frac{1}{M'^2}\Tr[W_{\mu}W^{\mu}]B_{\alpha\beta}{B}^{\alpha\beta}\,.
\end{align}
Here $W$ and $B$ correspond to the dark gauge field and the SM hypercharge field, and $M'$ is some (high) mass scale in the UV. Similarly, if there are some heavy Dirac fermions charged under both the dark $SU(2)$ and $U(1)_{Y}$ of the SM, it may generate Euler-Heisenberg terms:
\begin{align}
    \Delta\mathcal{L} \sim &\, \frac{\alpha_{Y}\alpha_{d}}{M_f^4}\Tr[G_{\mu\nu}\tilde{G}^{\mu\nu}]B_{\alpha\beta}\tilde{B}^{\alpha\beta}\nonumber\\
    &+ \frac{\alpha_{Y}\alpha_{d}}{M_f^4}\Tr[G_{\mu\nu}G^{\mu\nu}]B_{\alpha\beta}B^{\alpha\beta},
\end{align}
in the low energy effective theory. Apart from non-renormalizable operators, there could be renormalizable ones as well. Examples include the Higgs portal where the SM Higgs couples to the dark SU($2$) Higgs via a dimension 4 operator
\begin{align}
    \Delta\mathcal{L} \sim \lambda_{h\varphi}\,\Phi^{\dagger}\Phi\,H^{\dagger}H\,,
\end{align}
with the coupling $\lambda_{h\varphi} \lesssim 0.01$ from current observational bounds~\cite{Arcadi:2019lka} (for the dark Higgs mass $M_{\varphi} \lesssim $ GeV). 

Such interactions, although suppressed by tiny effective couplings, can have interesting phenomenological implications owing to resonant Higgs and/or photon production, similar to the resonant photon production in axion-photon case~\cite{Hertzberg:2018zte,Amin:2020vja,Amin:2021tnq}. Albeit there being an important distinction due to spin and/or iso-spin.\\

\noindent\textbf{Dark matter core vs radius relationship}: Observations of dwarf galaxies suggest that the dark matter central densities scale inversely with their radii, $\rho_{\mathrm{core}} \propto r_s^{-1}$~\cite{Deng:2018jjz,burkert2020fuzzy,2020ApJ...893...21S}. Due to the presence of repulsive self interactions, Chandrasekhar solitons have the scaling $\rho_{\mathrm{core}} \propto M_s$, i.e. it doesn't depend on the radii. Since the HYM sector can house both classes of solitons (attractive or repulsive self interactions within), depending upon the masses $m, M_{\varphi}$ and the gauge coupling $g$, it may provide for a better fit to the observed data (over the usual $\rho_{\mathrm{core}} \propto r_s^{-4}$ scaling for solitons supported primarily by a balance between gravity and gradient pressure).\\

For Chandrasekhar solitons that are at-least on the order of a few kpc and hence more relevant for such dwarf galaxy halo observations, we need $m \lesssim 0.1$ eV (c.f.~\eqref{eq:rs_bound}).
While on one hand this would require somewhat smaller gauge couplings $g \lesssim 10^{-6}$ (c.f.~\eqref{eq:g_bound} due to Bullet cluster constraints), on the other it would make such solitons tremendously long lived. For such values of $m$, Chandrasekhar solitons that are at-least as long lived as the age of the Universe, can be as massive as $10^{5}-10^{6}\, M_{\odot}$ and hence be relevant for the cores of dark matter halos.\\

Note that this is in stark contrast to the usual scenario of having solitonic like cores in DM halos, where the mass $m$ of the DM particle is in the fuzzy regime. Interestingly, $m \lesssim 0.1$ eV falls within the preferred mass range for the QCD axion. For the production mechanism of~\cite{Agrawal:2018vin} then, one can have the QCD axion as the misaligned scalar itself.\\

A careful investigation of the production mechanism of such a HYM sector, including the formation and properties of DM halos for different values of $m$, $M_{\varphi}$ and $g$, is needed. See~\cite{Gorghetto:2022sue,Amin:2022pzv} for a recent analysis for a single vector field without any self-interactions. It remains to be seen whether having all possible soliton scalings can introduce enough scatter into the core density vs radius relation.\\

\noindent\textbf{Diversity of dark matter mass and larger gauge groups}: In this work we focused mainly on the case of SU($2$), with Higgs in the fundamental representation. However there are many other possibilities that are worth exploring. In the adjoint representation, and especially for SU($n \geq 3$), we can admit a diverse set of mass values for the different vector fields depending upon the number of Higgs fields and their vev matrices. This may further help to address the observed core mass and radii values at different hierarchical scales. For instance see~\cite{Luu:2018afg} for a set of scalar fields with different masses (in the regime $m \sim 10^{-20}-10^{-21}$ eV).\\

\acknowledgments

I am grateful to Mustafa A. Amin for many useful discussions regarding solitons and oscillons, Mark Hertzberg regarding lifetimes and decay rates of solitons with different kinds of self-interactions, and Andrew Long for different production mechanisms of vector fields, including thermal production mechanisms in general. I would also like to thank them for their suggestions and comments on this draft. This work is supported in part by DOE-0000250746.

\bibliography{reference}

\end{document}